\documentclass[a4paper,12pt]{article}
\pdfoutput=1
\usepackage{lmodern}
\usepackage{jheppub}	
\usepackage{graphicx}				
\usepackage{amsmath}
\usepackage{mathtools}
\usepackage{amssymb}
\usepackage{booktabs}
\usepackage{comment}
\usepackage{dsfont}
\usepackage{amsthm}
\usepackage{mathrsfs}

\usepackage{multirow}
\usepackage{overpic}
\usepackage{caption}
\usepackage{subcaption}
\usepackage{array}
\usepackage{rotating}

\usepackage[T1]{fontenc} 

\usepackage{color}
\definecolor{dark-gray}{gray}{0.20}
\definecolor{gray}{gray}{0.30}
\definecolor{light-gray}{gray}{0.80}
\definecolor{dark-red}{rgb}{0.7,0,0}
\definecolor{dark-green}{rgb}{0.1,0.4,0}
\definecolor{dark-blue}{rgb}{0.3,0.3,0.7}
\definecolor{light-blue}{rgb}{0.8,0.8,1}

\usepackage{hyperref}
\hypersetup{
	colorlinks=true,
	linkcolor=dark-blue,
	citecolor=dark-red,
	urlcolor=dark-green,
	linktoc=page
}

\title{\boldmath State-independent Black Hole Interiors from the Crossed Product}

\author[a]{Chethan Krishnan,}
\author[b]{Vyshnav Mohan}

\affiliation[a]{Center for High Energy Physics,\\
Indian Institute of Science, Bangalore 560012, India.}
\affiliation[b]{Science Institute,University of Iceland \\Dunhaga 3, 107 Reykjav\'{i}k, Iceland.}

\emailAdd{chethan.krishnan@gmail.com}
\emailAdd{vyshnav.vijay.mohan@gmail.com}

\abstract{Opinion is divided about the nature of state dependence in the black hole interior. Some argue that it is a necessary feature, while others argue it is a bug. In this paper, we consider the extended half-sided modular translation $U(s_0)$ (with $s_0 > 0$) of Leutheusser and Liu that takes us inside the horizon. We note that we can use this operator to construct a modular Hamiltonian $H$ and a conjugation $J$ on the infalling time-evolved wedges. The original thermofield double translates to a new cyclic and separating vector in the shifted algebra. We use these objects and the Connes cocycle to  repeat Witten's crossed product construction in this new setting, and to  obtain a Type II$_\infty$ algebra that is independent of the various choices, in particular that of the cyclic separating vector. Our emergent times are implicitly boundary-dressed. But if one admits an ``extra'' observer in the interior, we argue that the (state-independent) algebra can be Type I or Type II$_1$ instead of Type II$_\infty$, depending on whether the observer's light cone contains an entire Cauchy slice or not. Along with these general considerations, we  present some specific calculations in the setting of the Poincare BTZ black hole. We identify a generalization of modular translations in BTZ-Kruskal coordinates that is  pointwise (as opposed to non-local) and is analytically tractable, exploiting a connection with the covering AdS-space.  These evolutions can reach the singularity.}

\makeatletter
\gdef\@fpheader{}
\makeatother

  {\itshape}
  {}

\begin{document} 
\newcommand{\beq}{\begin{equation}}
\newcommand{\eeq}{\end{equation}}
\newcommand{\bea}{\begin{eqnarray}}
\newcommand{\eea}{\end{eqnarray}}
\newcommand{\beas}{\begin{eqnarray*}}
\newcommand{\eeas}{\end{eqnarray*}}
\newcommand{\defi}{\stackrel{\rm def}{=}}
\newcommand{\non}{\nonumber}
\newcommand{\bquo}{\begin{quote}}
\newcommand{\enqu}{\end{quote}}
\renewcommand{\(}{\begin{equation}}
\renewcommand{\)}{\end{equation}}
\def \eqn#1#2{\begin{equation}#2\label{#1}\end{equation}}

\newtheorem*{remark}{Proposal}

\def\e{\epsilon}
\def\IZ{{\mathbb Z}}
\def\IR{{\mathbb R}}
\def\IC{{\mathbb C}}
\def\IQ{{\mathbb Q}}
\def\IH{{\mathbb H}}
\def\de{\partial}
\def\Tr{ \hbox{\rm Tr}}
\def\H{ \hbox{\rm H}}
\def\HE{ \hbox{$\rm H^{even}$}}
\def\HO{ \hbox{$\rm H^{odd}$}}
\def\K{ \hbox{\rm K}}
\def\Im{ \hbox{\rm Im}}
\def\Ker{ \hbox{\rm Ker}}
\def\const{\hbox {\rm const.}}
\def\o{\over}
\def\im{\hbox{\rm Im}}
\def\re{\hbox{\rm Re}}
\def\bra{\langle}\def\ket{\rangle}
\def\Arg{\hbox {\rm Arg}}
\def\Re{\hbox {\rm Re}}
\def\Im{\hbox {\rm Im}}
\def\exo{\hbox {\rm exp}}
\def\diag{\hbox{\rm diag}}
\def\longvert{{\rule[-2mm]{0.1mm}{7mm}}\,}
\def\a{\alpha}
\def\dag{{}^{\dagger}}
\def\tq{{\widetilde q}}
\def\p{{}^{\prime}}
\def\W{W}
\def\N{{\cal N}}
\def\hsp{,\hspace{.7cm}}

\def\br{\nonumber}
\def\IZ{{\mathbb Z}}
\def\IR{{\mathbb R}}
\def\IC{{\mathbb C}}
\def\IQ{{\mathbb Q}}
\def\IP{{\mathbb P}}
\def \eqn#1#2{\begin{equation}#2\label{#1}\end{equation}}

\newcommand{\C}{\ensuremath{\mathbb C}}
\newcommand{\Z}{\ensuremath{\mathbb Z}}
\newcommand{\R}{\ensuremath{\mathbb R}}
\newcommand{\rp}{\ensuremath{\mathbb {RP}}}
\newcommand{\cp}{\ensuremath{\mathbb {CP}}}
\newcommand{\vac}{\ensuremath{|0\rangle}}
\newcommand{\vact}{\ensuremath{|00\rangle}                    }
\newcommand{\oc}{\ensuremath{\overline{c}}}
\newcommand{\psizero}{\psi_{0}}
\newcommand{\phizero}{\phi_{0}}
\newcommand{\hzero}{h_{0}}
\newcommand{\psiin}{\psi_{\rh}}
\newcommand{\phiin}{\phi_{\rh}}
\newcommand{\hin}{h_{\rh}}
\newcommand{\rh}{r_{h}}
\newcommand{\rb}{r_{b}}
\newcommand{\psibnd}{\psi_{0}^{b}}
\newcommand{\psibndp}{\psi_{1}^{b}}
\newcommand{\phibnd}{\phi_{0}^{b}}
\newcommand{\phibndp}{\phi_{1}^{b}}
\newcommand{\gbnd}{g_{0}^{b}}
\newcommand{\hbnd}{h_{0}^{b}}
\newcommand{\zh}{z_{h}}
\newcommand{\zb}{z_{b}}
\newcommand{\man}{\mathcal{M}}
\newcommand{\hbr}{\bar{h}}
\newcommand{\tbr}{\bar{t}}

\maketitle
\flushbottom












\section{Introduction and Conclusions}

The AdS/CFT correspondence \cite{Maldacena:1997re, Witten:1998qj} has provided us with an explicit theory of quantum gravity in $\mathcal{N}=4$ super Yang-Mills theory. This has shifted the central problem of quantum gravity from efforts to quantize (bulk local) Einstein gravity, to that of understanding the emergence of bulk locality from the hologram. 

The reconstruction of a bulk local description outside a black hole (or when there is no black hole) from the boundary CFT is relatively uneventful \cite{HKLL2}. But going beyond the horizon is challenging \cite{Kraus, Fidkowski, FestucciaLiu, HKLL3, HKLL4, Papadodimas:2013jku} because the causal structure has to be emergent. In thought-provoking papers, Papadodimas and Raju have exploited the connection with the eternal black hole thermofield double state \cite{Israel:1976ur, Maldacena:2001kr} to reconstruct the interior, and argued that it is necessarily state-dependent \cite{Papadodimas:2013jku, Papadodimas:2015jra}. But others have argued that state-dependence gives rise to ambiguities \cite{Harlow} and interpretational challenges \cite{MarolfPolchinski2}. In this paper, 
we will argue that in perturbation theory around the large-$N$ limit, the interior reconstruction can in fact be made naturally state {\it in}dependent\footnote{Our discussions in this paper are to first order in $1/N^2$, but we expect the conceptual picture to remain valid to all orders. Discussing algebras over the ring ${\mathbb C}[[1/N]]$ involves some  subtleties \cite{CPW}, but we expect them to not be insurmountable. However, we do expect conceptually new ingredients at finite $N$, which we will comment on.}. At finite-$N$ the exterior algebras are expected to be Type I \cite{Leutheusser:2021frk, Witten:2021unn} and therefore the interior reconstruction suggested in \cite{Papadodimas:2013jku, Papadodimas:2015jra} (which postulates a {\it bulk} Kruskal time), cannot be exact. So we believe our results in this paper are sufficient to address the current state of the art. Our results sharpen the claims of \cite{Papadodimas:2013jku, Papadodimas:2015jra} about the possibility of interior reconstruction -- by showing that in the perturbative large-$N$ limit, state-dependence is not a problem. But our results are also strongly suggestive that at finite-$N$ and late times, interior reconstructions that rely on the thermal state are unlikely to be trustable\footnote{This has important consequences for arguments based of TFD states that claim to resolve firewalls. We will discuss the relevant timescales as well as the status of firewalls, in various parts of the paper. See in particular Section \ref{intro-sub}.}.


There are two key recent developments that we will take advantage of in arriving at our results. The first is the construction of {\it emergent} bulk times in holographic duality in the large-$N$ limit, due to Leutheusser and Liu \cite{Leutheusser:2021frk, Leutheusser:2021qhd}. This approach is appealing because it is a boundary-construction of bulk time and casual structure. These authors identify an extension of half-sided modular translations that takes us inside the horizon, exploiting the fact that the bulk algebra is of Type III$_1$ in the large-$N$ limit. Modular translations exist only in Type III$_1$ von Neumann algebras, and the construction of \cite{Leutheusser:2021frk, Leutheusser:2021qhd} is manifestly dependent on the thermofield double. Our main observation is that another recently noted ingredient, this time due to Witten \cite{Witten:2021unn}\footnote{For recent applications of related ideas, see eg. \cite{Leigh,Gesteau}.}, can be used to cure this problem and construct a state-independent emergent Type II$_\infty$ algebra in the interior. Our key strategy will be to use the extended modular translation to evolve the thermofield double state to a new state in the new algebra. This state can be shown to be cyclic and separating. We observe that the modular Hamiltonian of this new state, together with the (shared) center of the left and right algebras allows us to repeat Witten's crossed product construction. This is sufficient to argue that state-dependence is an artifact of the strict $N=\infty$ limit and that it is absent once the perturbative $1/N^2$ corrections are incorporated. (The precise connection between these technical results and the intuitive notion(s) of state-dependence are clarified later in this Introduction as well as elsewhere.)

The argument above (see Section \ref{proof} for details) will show that the resulting algebra is independent of the {\it emergent} cyclic separating vector in the emergent time-evolved wedges. But this raises another question. Our observation does not immediately show that the construction is independent of the {\it initial} cyclic separating vector that we started with. To demonstrate that it is independent of this initial state, we will use the Connes cocycle (see Section \ref{Connes}). This object has made an appearance in the physics literature (see eg. \cite{LashkariConnes}). Composing the unitary relating two cyclic separating vectors with the modular translation, we will be able to show that there is a {\it unique} conjugacy between algebras built from distinct cyclic separating vectors (which is what one understands by state-independence \cite{Witten:2021unn}). 

The isomorphisms between the interior algebras have a natural interpretation in terms of their representations. We should consider the algebras themselves as state-independent entities. However, when we want to construct explicit representations of the algebra, we need to pick a particular state. These representations are then connected through a unique conjugation.


We will also make some comments (see Section \ref{subalgebra}) about another choice involved in the construction of modular translations, which is somewhat tangential to the main thrust of this paper, but may be of broader conceptual interest. To construct a modular translation, we need to choose a sub-region (denoted $\mathcal{N}$ in \cite{Leutheusser:2021frk}) of the original wedge (denoted $\mathcal{M}$ in \cite{Leutheusser:2021frk}). We will be able to construct an equivalence relation between operators at distinct exterior points, when there exists modular translations built from two different subregions $\mathcal{N}$ and $\tilde{\mathcal{N}}$, that move these points to the same bulk point in the interior. This equivalence can be constructed by modular translating forward with the translation built from $\mathcal{N}$ and then backward using a translation built from $\tilde{\mathcal{N}}$. The resulting unitary connects the two pre-image points in the exterior. We suspect that this unitary should be viewed as a (emergent) symmetry of the quantum field theory in the wedge $\mathcal{M}$. It is easy to see for simple choice of subalgebras in the ordinary Minkowski/Rindler setting, that this construction leads us to the re-discovery of ordinary Poincare translations. It will be very interesting to explore the nature of emergent modular symmetries in general spacetimes in a holographic setting.

In Section \ref{more} we will outline how the construction of \cite{Papadodimas:2013jku,Papadodimas:2015jra} is related to that of \cite{Leutheusser:2021frk}. The goal is to clarify the distinctions and connections between the two approaches and to argue that in the large-$N$ limit, bulk reconstruction should be well-defined without the possible issue of state-dependence for thermal states far from typicality. These discussions are closely related to the question of firewalls \cite{Mathur, AMPS} at late times \cite{MarolfPolchinski1, Susskind}, which we also make comments on.

The discussions above are self-contained and a reader who is only interested in the general arguments can skip over Section \ref{BTZ}. Section \ref{BTZ} is somewhat technical and discusses modular translations of the Poincare-BTZ black hole. One of our goals here is to show that one can construct a fully analytic and pointwise (as opposed to non-local) example of a generalized modular translation that can take us inside the horizon of BTZ. To contrast, the construction across the BTZ horizon in \cite{Leutheusser:2021frk} used a near horizon approximation. Our translation instead works directly with the full BTZ metric, and exploits the fact that BTZ is locally AdS$_3$ to make a connection with AdS-Rindler. The modular translations in the latter are known \cite{Leutheusser:2021frk}, and therefore we can adapt them. The price we pay is that the resulting expressions are somewhat ugly -- but we explicitly check that they take us inside the horizon. In fact, because we know them quite explicitly and because they are valid beyond the near-horizon Rindler approximation, we can show that these translations can take us all the way to singularity. 

In Section \ref{infaller},  we briefly comment on the possibility of adding an (infalling) observer in the interior as a mechanism for constructing background independent algebras in quantum gravity, as has recently been suggested \cite{Witten:2023qsv, Witten:2023xze}. This is logically somewhat distinct from our discussion in Section \ref{proof}, which can be viewed as the emergence of the interior in a boundary-dressed setting as opposed to an observer-dressed setting. If one simply {\it assumes} that the interior observer is a well-defined entity in the code subspace of the bulk EFT in the black hole\footnote{That there is a self-consistent description of an observer interacting with the rest of the system in the bulk EFT code subspace seems to be the premise of the arguments in \cite{Chandrasekaran:2022cip, Witten:2023qsv, Witten:2023xze}. In the black hole interior, it would be worth scrutinizing this assumption more intensely than in other contexts. The resulting algebras we find here seem natural, so perhaps it is indeed a reasonable assumption here as well.}, we can instead investigate the nature of the emergent algebras in this modified setting. Interestingly, this leads to some interesting distinctions from some of the other discussions of observer-dressed algebras. We find two qualitatively different classes of timelike curves for the observers in a Kruskal BTZ geometry -- (a) those that start at the white hole singularity and end at the future singularity, and (b) those that start from one of the boundaries and end in the future singularity. The latter class of observers can be called infallers. We will argue that the natural algebras one can associate in the two cases are different. In the former, we are led to a Type I algebra, and in the latter, we find Type II$_1$. 
We will find that these results are natural both from the bulk as well as the boundary. 

\subsection{On State-independence}\label{intro-sub}

The phrase state-(in)dependence can mean different things in different settings, so let us clarify some of the relevant ideas before closing the Introduction. The goal is to give a physical understanding of our results and their origin, as well as their consequences.

The first notion of state-independence is somewhat trivial and refers to the fact that one can always find an isomorphism between the states (and operators) of two Hilbert spaces (as long as their dimensions are the same). We will describe this in more detail in Section \ref{WvsS}, and will emphasize that the isomorphism we construct is canonical in the sense of the crossed product construction of \cite{Witten:2021unn}. This shows that the interior reconstructions are as state-independent as the exterior reconstructions are, before the Page time (more on the timescales below). 

A more substantive distinction needs to be made between the state-dependence we discuss and background dependence. The latter was discussed in \cite{Witten:2023xze} recently. The idea here is that in full quantum gravity, the algebras one deals with are presumably independent of specific choices of spacetime, because the spacetime is specified by the state and we would like the algebras to be independent of states in quantum gravity. But when we are working with quantum field theory in curved spacetime, spacetimes are generally viewed as given a-priori and the state dependence is something we can address at the level of given spacetimes and subregions/algebras. The state-dependence we are concerned with in our paper has to do with states in the given background. The choice of the spacetime goes e.g., into the background subtraction of the black hole mass that we do in eg., \eqref{backgroundsubtraction}. More generally, it arises also in defining the ``subtracted'' single-trace operators (where one subtracts the expectation values from single-trace operators) to define a well-defined large-$N$ limit \cite{Witten:2021unn}. We feel that this is a step that could benefit from more scrutiny than what it has received so far. Our approach in this paper is less ambitious. It is useful perturbatively around the large-$N$ limit and therefore should be relatively non-controversial, but at finite $N$ one will have to worry about background independence of the stronger kind. 

The above distinction is not merely a technicality. If one does not make a distinction between non-perturbative vs. perturbative $1/N$ effects, then one can consider timescales that are $\mathcal{O}(N^\#)$ or $e^{\mathcal{O}(N)}$ in the eternal black hole spacetime\footnote{We will call these the Page time and the typicality time, respectively. Here $\#$ is some positive number whose precise value depends on details like the dimensionality of the spacetime, and is unimportant.}. This is the attitude of \cite{Papadodimas:2013jku} who do not distinguish between finite and infinite $N$ and postulate a smooth interior time coordinate even at finite $N$\footnote{Their construction also has some other differences compared to \cite{Leutheusser:2021frk} -- eg., they work with bulk wave equations more directly. These are unimportant distinctions for our purposes. The fact that interior operators are (at least manifestly) state-dependent is true for both constructions.}. On the contrary, we are working in a large-$N$ limit, where the backreaction of perturbative states is no longer possible\footnote{We emphasize that this is essential for the \cite{Leutheusser:2021frk, Leutheusser:2021qhd} construction to make sense, so we are {\em not} adding anything by hand. This is implicit in the fact that the black holes under consideration are, eg.,  above the Hawking-Page transition.}. This avoids the ambiguities suggested in \cite{Harlow} which arise because acting with a sufficient number of single trace operators can result in changing the background ``equilibrium'' state (at finite-$N$). This is  the simple conceptual reason why we are able to make the construction state-independent and unambiguous.

Let us phrase the above observations in a slightly different (and perhaps more insightful) way. We start by emphasizing that the state dependence we are concerned with is quite literally the dependence of the interior operators on the state in the thermofield double Hilbert space\footnote{See also our discussions in Sections \ref{more} and \ref{summary}.}. Now, constructions of interior operators that rely on the thermofield double \cite{Papadodimas:2013jku, Leutheusser:2021qhd}, are predicated on the reliability of the thermal (Hartle-Hawking) correlator. But it is crucial to remember that the Hartle-Hawking correlator with its exponential decay cannot be the whole story, beyond the Page time\footnote{In all these discussions $t=0$ is the insertion time of the first operator in the Hartle-Hawking two-point function. Let us also note that in order to talk about a finite Page time, we must necessarily work at finite $N$.}. This was emphasized in \cite{Maldacena:eternal}, and it has been noted since then in multiple contexts that one can expect changes in the qualitative behavior of the correlator after Page time\footnote{See eg., \cite{Cotler, Yang, Kaplan, Burman}. These papers note hints of discreteness of the spectrum (often via the tips of Wigenr semi-circle distributions) at around the Page time, in SYK model, JT-gravity, (Virasoro vacuum blocks in) large-$c$ 2D CFT and bulk models of black hole microstates. We refer the reader to Section 6 of \cite{Burman} for a related discussion.}. So by noticing that state-{\it in}dependence can be accomplished using (extended) modular translations (which exist only when the Hartle-Hawking correlator is meaningful\footnote{Modular translations exist only when the algebra is Type III. That the spectrum of the Hartle-Hawking correlator has support on the entire real line, was one of the arguments in  \cite{Leutheusser:2021qhd} for the emergence of Type III.}), our results demonstrate that the ambiguities related to state-dependence can be avoided at pre-Page times. This is also a suggestion that the ambiguities that arise (which must necessarily be at post-Page timescales) are due to an incorrect application of the {\em exact}  thermal correlator (and the associated thermofield double) at very late times where they are no longer valid\footnote{Our construction (and indeed that of \cite{Leutheusser:2021qhd}), being at large-$N$, does not directly deal with questions about firewalls-vs-smoothness at late times in eternal AdS black holes. But it raises significant questions about the validity of thermofield double-based constructions which claim to remove firewalls even at very long timescales.}. This also clarifies the regime of validity of HKLL based reconstructions of the interior -- they also rely on the validity of the thermal correlator to do analytic continuations into the interior.

To summarize -- state dependence is present in both \cite{Papadodimas:2013jku} as well as \cite{Leutheusser:2021frk} constructions. It is not directly related to the dynamical emergence (or not) of interior time, it is about the fact that interior operators depend on the chosen state. Both constructions potentially suffer from the ambiguity problem raised in \cite{Harlow}. But the key advantage of the \cite{Leutheusser:2021frk} approach is that it shows where exactly this issue can be resolved, and what is causing the problem when there are ambiguities. It traces the origin of ambiguities to the breakdown of the Hartle-Hawking correlator beyond the Page time.

State/background independence of the emergent Type II crossed product algebras in the black hole exterior and in de Sitter space were discussed recently in \cite{Witten:2021unn, Chandrasekaran:2022cip}. In the context of discussions about the black hole interior, we have clarified and specialized these observations to make statements about state-independence. In particular, the boundary-dressed construction of Sections \ref{proof} and \ref{Connes} shows that state-independence of the black hole interior is automatic, when perturbative deviations from the strict large-$N$ limit are incorporated. The discussion in Section \ref{infaller} assumes the existence of an infalling/interior observer, and we find that the resulting algebras have natural features which may be useful in understanding the black hole singularity. Some further comments are made in the Discussion section \ref{disc}.  A reader who is only interested in an overview of our results and their implications can adopt a track through the paper that consists of the present section, the introductory page of Section \ref{proof}, Sections \ref{WvsS}, \ref{subalgebra}, \ref{more}, \ref{singularity}, \ref{infaller} and \ref{disc}.

\section{Modular Translations and Crossed Product in the Interior}\label{proof}

The crossed product construction goes back in the physics and mathematics literature to the 60's and 70's \cite{AharonovSusskind, Takesaki, Connes}. It was implemented recently in a form directly useful for us by Witten in the setting of eternal AdS-black holes in large-$N$ AdS/CFT \cite{Witten:2021unn}. The basic idea here is that in the background of an eternal black hole, there exists a shared central element for the algebras of fluctuations on both the left and right sides in the large-$N$ limit. The central element generates time translations in the boundary theories. There is a natural modular Hamiltonian one can associate to these algebras, which is associated to the thermo-field double (TFD) state, a cyclic separating vector. The trouble is that simply adjoining this element to the algebras results in evolutions that are dependent on the state. This is state-dependence\footnote{In previous black hole interior constructions like \cite{Papadodimas:2013jku} the bulk interior time coordinate was postulated, and the detailed construction was different from \cite{Leutheusser:2021qhd} and typically involved solutions of bulk wave equations. But we emphasize that the conceptual problems can ultimately be traced back again to state dependence -- the definition of  interior operators depended on the choice of the cyclic separating state, and this is what leads to ambiguities \cite{Harlow}.}. But the presence of the shared central element (which is something meaningful to talk about once you incorporate the perturbative $1/N^2$ corrections) gives us a different way of adjoining the modular Hamiltonian. This alternate operation, the crossed product, has the advantage that it is independent of the choice of the cyclic separating vector in a precise mathematical sense (see next section). 

The precise definition of the crossed product is presented in Appendix \ref{crossedproductrevsec} as well as later in this section in a more operational form. But for the moment, all we need to know is that the crossed product construction can be done once we have a modular Hamiltonian and a (shared) center. For the eternal black hole exterior algebras, the modular Hamiltonian is simply the one associated to the TFD state. Witten's crucial observation \cite{Witten:2021unn} was that in perturbation theory around the large-$N$ limit, one also has a natural central element which he called $U$. Before the crossed product, the  left and right algebras are
\bea
\mathcal{A}^0_L=\mathcal{A}_L \otimes \mathcal{A}_U, \ \ \mathcal{A}^0_R=\mathcal{A}_R \otimes \mathcal{A}_U
\eea
Here $\mathcal{A}_{L/R}$ stands for the algebras of small fluctuations around the left and right wedges of a black hole of fixed mass. The central element generates time translations in the bulk. The central element $U$ is accessible to both the left and the right. Together with the modular Hamiltonian, $U$ allows us to do the crossed product construction and make the exterior algebras state-independent.

The modular translations of Leutheusser and Liu are defined on the $\mathcal{A}_{L/R}$ algebras above, and lead to what we will call the infalling time evolved wedge algebras (see below). Our starting observation is that there is a natural modular Hamiltonian one can associate to these modular translated algebras, and therefore one can repeat the Witten crossed product construction for these algebras as well, using an infalling time evolved $U$ as the central element. The result is a Type II$_\infty$ algebra that is at least superficially state-independent. There will be a few subtleties in this somewhat quick statement, which we will iron out in the next section. In the rest of this section, we elaborate on the basic point. 

\subsection{Modular Translated Modular Hamiltonian}

Consider an eternal AdS black hole with two boundaries. Let us look at the boundary theories at large $N$. The Hilbert space of small fluctuations about the thermofield double $\Psi_{\text{TFD}}$ can be obtained by acting with single trace operators on it. The algebra of these single trace operators forms a von Neumann algebra of type III$_1$ above the Hawking-Page temperature $T>T_{HP}$ \cite{Leutheusser:2021frk,Furuya:2023fei}. The thermofield double state is a cyclic and separating vector w.r.t the algebra of these single trace operators. We can associate a modular operator $\Delta$ and a modular Hamiltonian $H = -\log{\Delta}$ to this state.

The single trace operators are dual to small fluctuations in the bulk. Let us denote the bulk operators restricted to the exterior right wedge of the black hole by $\mathcal{O}_R$. The large-$N$ limit of the boundary theory is dual to the $G_N \to 0$ limit in the bulk. Therefore, the algebra associated to the bulk operators, which we will denote by $\mathcal{A}_R$, simplify. Moreover, it is generated by free fields and forms a von Neumann algebra of type III$_1$ \cite{Leutheusser:2021frk}. We will denote the commutant of this algebra by $\mathcal{A}_L$. It was argued in \cite{Leutheusser:2021frk} that the half-sided modular translations, generated by $U(s)$ can be extended to $s>0$ for generalized free fields. With suitable choices, we can arrange these translations to take the operators across the horizon when $s>0$.

\begin{figure}
\centering
\includegraphics[width=1\linewidth]{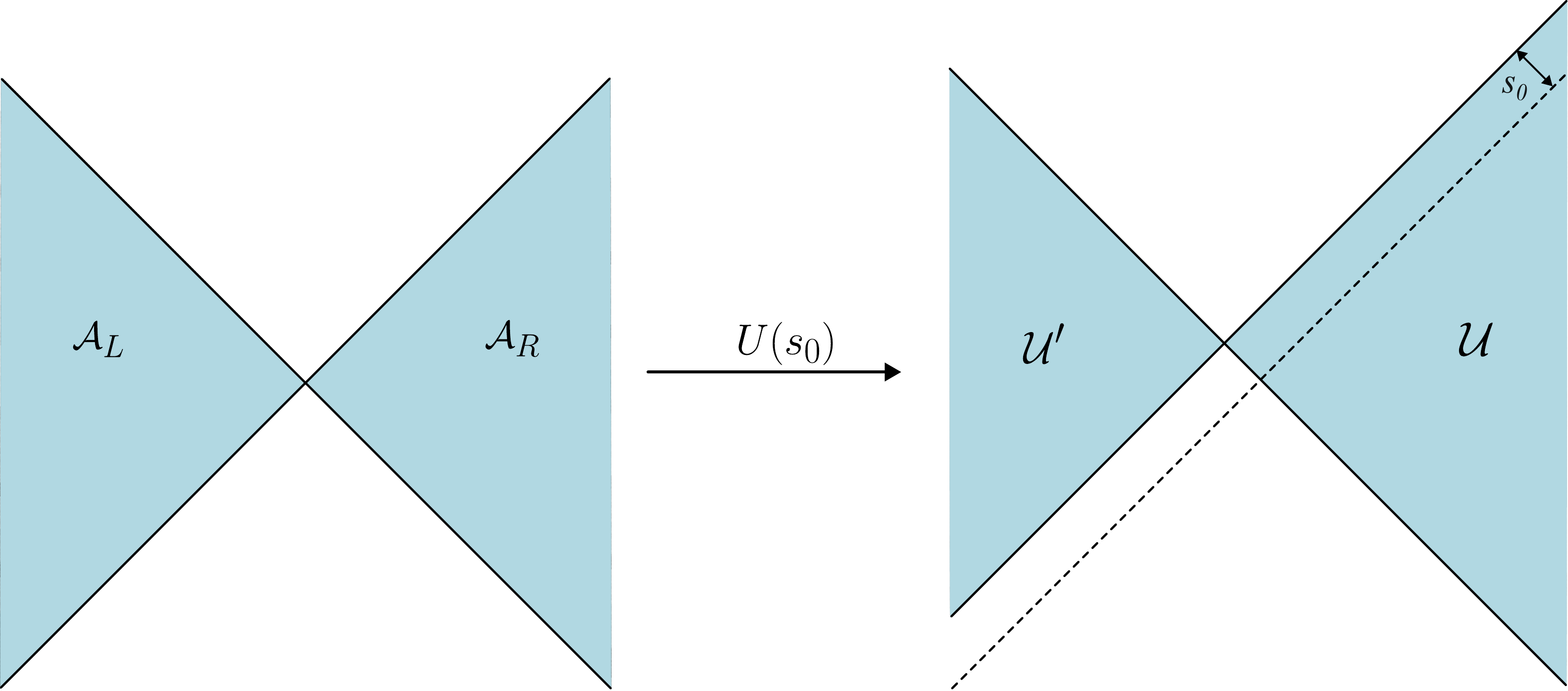}
\caption{Consider the two exterior regions of an eternal AdS black hole. In the large-$N$ limit, the algebra of operators associated to these wedges form type III$_1$ algebras $\mathcal{A}_{L,R}$. Under the evolution of $U(s_0)$, the right wedge algebra gets pushed inside the interior as shown in the figure. The dashed line denotes the location of the horizon.}
\label{shiftedfig}
\end{figure}

Let us define the following algebra of operators 
\bea
\mathcal{U} = \{U^{\dagger}(s_0)\mathcal{O}_RU(s_0) \ |  \ \forall \mathcal{O}_R \in \mathcal{A}_R  \} \quad \quad \text{for some} \ s_0 \in \left(0,\infty\right). \label{shiftedalgebradef}
\eea
The choice of $s_0$ here is arbitrary, but fixed. By varying $s_0$ we can go deeper into the interior (See figure \ref{shiftedfig}). The algebra $\mathcal{U}$ comprises of operators in the shifted wedge in figure \ref{shiftedfig} and depends on the details of $U(s_0)$. In the BTZ case that we will consider in Section \ref{BTZ}, $U(s_0)$ corresponds to a translations along a Poincare Rindler/BTZ coordinate $x^{-}$. In more general spacetimes, we can work in the near-horizon limit where all black holes have a Rindler region. This structure can be used to define a half-sided modular translation \cite{Leutheusser:2021frk}.

We claim that we can associate a new cyclic and separating vector to this algebra, given by
\bea
\tilde{\Psi}_{\text{TFD}} = U^{\dagger}(s_0) \Psi_{\text{TFD}}.\label{newtfdeq}
\eea
Associated to this vector is our new modular operator
\bea
\tilde{\Delta} = U^{\dagger}(s_0)\Delta U(s_0). \label{newmodularoperatoreq}
\eea
Now\footnote{The rest of the results in this subsection follow from standard facts regarding the Tomita operator, modular operator, and modular conjugation for algebras that have been conjugated by a unitary, and states to which the same unitary has been applied.}, let us demonstrate that $\tilde{\Delta}$ does, in fact, act as a modular operator for the algebra $\mathcal{U}$. To establish this,  we note that the modular Hamiltonian satisfies the KMS condition w.r.t its cyclic separating vector. In fact, this is a sufficiency condition for an operator to be a modular Hamiltonian. Using \eqref{newtfdeq},and \eqref{newmodularoperatoreq}, we arrive at
\bea
\bra \tilde{\Psi}_{\text{TFD}}| \tilde{b}^{\dagger}\tilde{\Delta} \tilde{a}|\tilde{\Psi}_{\text{TFD}} \ket = \bra \tilde{\Psi}_{\text{TFD}}| \tilde{a} \tilde{b}^{\dagger}|\tilde{\Psi}_{\text{TFD}} \ket
\eea
which is precisely the KMS condition for the new modular operator \cite{Sorce:2023fdx}. Moreover, it also straightforward to show that the new modular operator generates automorphisms on $\mathcal{U}$: 
\bea
\tilde{\Delta}^{it}\mathcal{U}\tilde{\Delta}^{-it} = \mathcal{U}.
\eea
We can also define a modular conjugation operator for the algebra:
\bea
\tilde{J} = U^{\dagger}(s_0)J U(s_0). \label{newconjugoperatoreq}
\eea
Using \eqref{newtfdeq} and the properties of $J$, it is easy to show that $\tilde{J}$ satisfies the following properties
\bea
\tilde{J}|\tilde{\Psi}_{\text{TFD}} \ket = |\tilde{\Psi}_{\text{TFD}} \ket, \quad \bra \tilde{\Psi}_{\text{TFD}}| \tilde{a}\left(\tilde{J} \tilde{a}\tilde{J}\right)|\tilde{\Psi}_{\text{TFD}} \ket\geq 0 \  \tilde{a}\in \mathcal{U}.
\eea
as expected \cite{Araki1974SomePO}. Now let us look at the commutant of $\mathcal{U}$. The elements of this algebra, which we will denote by $\mathcal{U}^{\prime}$, are given by the action of $\tilde{J}$:
\bea
\mathcal{U}^{\prime} = \{\tilde{J}\tilde{a}\tilde{J} \ | \  \forall \tilde{a} \in \mathcal{U}  \}.
\eea
The algebra $\mathcal{U}^{\prime}$ can be obtained by the action of $U(s_0)$ on the commutant of $\mathcal{A}_R$.
For an element $\tilde{a}^{\prime}$ of $\mathcal{U}^{\prime}$, we have
\bea
\begin{aligned}
\tilde{a}^{\prime} = \tilde{J}\tilde{a}\tilde{J} &=\left(U^{\dagger}(s_0)J U(s_0)\right)\left(U^{\dagger}(s_0)a U(s_0)\right)\left(U^{\dagger}(s_0)J U(s_0)\right)=U^{\dagger}(s_0)a^{\prime} U(s_0) \hspace{0.2cm}
\end{aligned}
\eea
where we have $JaJ=a^{\prime} \in \mathcal{A}_L$.
This statement is precisely what we expect from the shifting of wedges (See figure \ref{shiftedfig}). The action of the operator $U(s_0)$ on the elements of $\mathcal{A}_L$ reduces to a half-sided modular translation \cite{Leutheusser:2021frk} and provides an additional check of the shifted wedge construction.

Therefore, the existence of $U(s)$ allows us to define a new von Neumann algebra $\mathcal{U}$, which has support both in the interior and the exterior of the black hole. In fact, we have constructed the quadruple - $(\mathcal{U}, \tilde{\Psi}_{\text{TFD}}, \tilde{\Delta},\tilde{J})$. One can also phrase this discussion in the language of the Tomita operator (of the new wedge) which is related to these entities. We do that in Appendix \ref{Tomita}.

\subsection{Mechanics of the Crossed Product}\label{mech}

The crossed product is a simple construction, and we provide a review of that in Appendix \ref{crossedproductrevsec}. We will follow the same notation here. Let us construct a crossed product algebra by adjoining $T+X$ to the shifted wedge algebra $\mathcal{U}$. A natural choice for $T$ is 
\bea
T = \frac{\tilde{H}}{\beta N}
\eea
where $\tilde{H} = -\log{\tilde{\Delta}}$ is the new modular Hamiltonian and $\beta$ is the inverse temperature of the black hole. We will choose the central operator to be
\bea
X = U^{\dagger}(s_0)\left(\frac{H^{\prime}_L}{N}\right) U(s_0),\label{Xoperatoreq}
\eea
Here $H^{\prime}_L$ is the subtracted left CFT Hamiltonian
\bea
H^{\prime}_L =H_L-\bra H_L \ket. \label{backgroundsubtraction}
\eea
The modular Hamiltonian defined by 
\bea
\beta H = H^{\prime}_R-H^{\prime}_L  \label{modularHCFTHeq}
\eea
is well-defined in the strict large-$N$ limit. So $H/N \rightarrow 0$ at large-$N$, which means that replacing $H^{\prime}_L/N$ with $H^{\prime}_R/N$ in \eqref{Xoperatoreq} as (central) element of the algebra is equivalent. The center was therefore denoted by $U$ in \cite{Witten:2021unn}, as well as in the introductory part of this section. Let us also note that $\bra H_L \ket$ is essentially the mass of the black hole, which in the large-$N$ limit is essentially infinite. This has the consequence that $H^{\prime}_L$ is unbounded from below which is responsible for the algebra being Type II$_\infty$ and not Type II$_1$.  

Now let us explain the motivation behind the choice \eqref{Xoperatoreq}. In \cite{Witten:2021unn}, the operator $X$ was chosen to be  (roughly speaking) the Hamiltonian of the left CFT so that the operator $T+X$ can be effectively thought to be generating time translations in the right wedge. Perhaps a bit surprisingly\footnote{Note that the time evolution in the interior is not directly accessible from the boundary. The following calculation can be viewed as a direct demonstration that the time evolutions in the interior are implicitly constructed from those in the original wedges.}, this holds true in our construction as well for the shifted right wedge. To see this, consider the commutator of an $\tilde{a} \in\mathcal{U}$ with $T+X$:
\bea
\left[T+X,\tilde{a}\right] =\left[T,\tilde{a}\right]+\left[X,\tilde{a}\right]
\eea
Let us look at the second term. We have
\bea
\begin{aligned}
\left[X,\tilde{a}\right] &= \left(U^{\dagger}(s_0)\left(\frac{H^{\prime}_L}{N}\right) U(s_0)\right)\left(U^{\dagger}(s_0)a U(s_0)\right)\\
&\hspace{4cm}-\left(U^{\dagger}(s_0)a U(s_0)\right)\left(U^{\dagger}(s_0)\left(\frac{H^{\prime}_L}{N}\right) U(s_0)\right)\\
&=0,
\end{aligned}
\eea
since $H^{\prime}_L$ commutes with the operators in the right wedge. This gives us
\bea
\begin{aligned}
\left[T+X,\tilde{a}\right] &= \left[\frac{\tilde{H}}{\beta N},\tilde{a}\right]\\
&=\frac{1}{\beta N}\bigg{[}\left(U^{\dagger}(s_0)H U(s_0)\right)\left(U^{\dagger}(s_0)a U(s_0)\right)\\
&\hspace{4cm}-\left(U^{\dagger}(s_0)a U(s_0)\right)\left(U^{\dagger}(s_0)H U(s_0)\right)\bigg{]} \\
\end{aligned}
\eea
where we have used
\bea
P = U^{\dagger}(s_0)Q U(s_0) \quad \implies \log{P} = U^{\dagger}(s_0)\left(\log{Q} \right)U(s_0).
\eea
to rewrite the new modular Hamiltonian. Simplifying the expression, we find
\bea
\begin{aligned}
\left[T+X,\tilde{a}\right] &= U^{\dagger}(s_0)\left[\frac{H}{\beta N},a\right]U(s_0)\\
&= U^{\dagger}(s_0)\left(-\frac{i}{N}\frac{\partial a}{\partial t}\right)U(s_0)\\
&= -\frac{i}{N}\frac{\partial }{\partial t}\left(U^{\dagger}(s_0)aU(s_0)\right)\\
&= -\frac{i}{N}\frac{\partial \tilde{a}}{\partial t}.
\end{aligned}
\eea
In the second equality, we have used \eqref{modularHCFTHeq} and the discussion above eqn. (3.2) in \cite{Witten:2021unn}. Therefore, we have demonstrated that $T+X$ generates the time translations of the operators in $\mathcal{U}$ -- which is precisely what the right CFT Hamiltonian did in \cite{Witten:2021unn}. This motivates us to adjoin the operator $T+X$ to the shifted wedge algebra through a crossed product construction.

The left-wedge discussion goes parallel to the left-wedge discussions in \cite{Witten:2021unn}, via a conjugation using $\Pi = -i \frac{d}{dU}$. The only difference is that the modular Hamiltonian that we use is the shifted one; we will not repeat the discussion here.

\section{Proof of State Independence via Connes Cocycle}\label{Connes}

Up to a unique equivalence, the crossed product of an algebra by its modular automorphism group is independent of its cyclic and separating vector. This statement can be proved using Connes cocycle \cite{connescocycle}. Therefore, the crossed product algebra $\mathcal{U} \rtimes \mathbb{R}$ obtained by adjoining the modular Hamiltonian $\tilde{\Delta}$ and the operator $X$ is independent of the cyclic and separating vector $\tilde{\Psi}_{\text{TFD}}$. However, it remains to be shown that our crossed product algebra is independent of the original state $\Psi_{\text{TFD}}$. Since the modular translation $U(s_0)$ implicitly depends on this choice, one might naively expect the shifted wedge algebra $\mathcal{U}$ to inherit this dependence. In this section, we will show that our crossed product algebra is also independent of the original cyclic separating vector.

Let us start by recalling the definition of the Tomita operator. Consider an algebra $\mathcal{A}$ acting on a Hilbert space $\mathcal{H}$. If $\Omega$ is a cyclic separating vector, the Tomita operator is given by
\bea
S_{\Omega} a \Omega = a^{\dagger} \Omega \quad \quad a \in \mathcal{A}. \label{tomitaeq}
\eea
The modular operator can be defined in terms of $S_{\Omega}$ as follows\footnote{In this section we will denote the (modular) translations associated to the vectors $\Omega$ and $\Psi$ as $U_{\Omega,\Psi}(s)$ and the modular Hamiltonians as $\Delta_{\Omega,\Psi}$. The corresponding infalling time-evolved modular Hamiltonians, see \eqref{newmodularoperatoreq}, will also have tildes. We will also use the definitions $X_{\Omega,\Psi} \equiv U^\dagger_{\Omega,\Psi}(s_0)\frac{H_L'}{N}U_{\Omega,\Psi}(s_0)$.}
\bea
\Delta_{\Omega} = S_{\Omega}^{\dagger}S_{\Omega}
\eea
If $\Psi$ is another cyclic separating vector, then we can define a relative Tomita operator and relative modular operator:
\bea
S_{\Psi|\Omega} a \Omega = a^{\dagger} \Psi \quad \quad \quad \quad \Delta_{\Psi|\Omega} = S_{\Psi|\Omega}^{\dagger}S_{\Psi|\Omega} \label{reltomitaeq}
\eea
Using these operators, one defines the Connes cocycle as
\bea
u_{\Psi \mid \Omega}(t)=\Delta_{\Psi \mid \Omega}^{\mathrm{i} t} \Delta_{\Omega}^{-\mathrm{i} t}=\Delta_{\Psi}^{\mathrm{i} t} \Delta_{\Omega \mid \Psi}^{-\mathrm{i} t}. \label{cocycleeq}
\eea
Connes cocycle satisfies numerous useful properties \cite{Witten:2018zxz,Witten:2021unn}. In particular, we can show that $u_{\Psi \mid \Omega}(t) \in \mathcal{A}$. It also maps modular flows under the change of cyclic separating vectors:\\
\noindent
{\bf Connes cocycle.} \begin{it}
Let $\Omega$ be a cyclic separating vector of the algebra $\mathcal{A}$. If $\Delta_{\Omega}$ is its modular operator, we can associate a modular flow, defined by $\sigma^{\Omega}_t(a) = \Delta_{\Omega}^{it}a\Delta_{\Omega}^{-it}, a \in \mathcal{M}$. If $\Psi$ is another cyclic separating vector, we have \cite{connescocycle}
\bea
\sigma^{\Psi}_t (a)= u_{\Psi \mid \Omega}(t)\sigma^{\Omega}_t(a)u_{\Psi \mid \Omega}^{-1}(t) \quad \quad \forall t \in \mathbb{R},\label{cocyclefloweq}
\eea
\end{it}
If $\mathcal{A}^{\prime}$ is the commutant of $\mathcal{A}$, then we can define operators $S^{\prime}_{\Omega}$ and $S^{\prime}_{\Psi|\Omega}$ using equations \eqref{tomitaeq} and \eqref{reltomitaeq} by replacing $a$ with an element $a^{\prime} \in \mathcal{A}^{\prime}$. Furthermore, we can define a Connes cocycle for the commutant, and it can be expressed using the modular operators of $\mathcal{A}$ as follows \cite{Witten:2021unn}:
\bea
u_{\Psi \mid \Omega}^{\prime}(t)=\Delta_{\Omega \mid \Psi}^{-\mathrm{i} t} \Delta_{\Omega}^{\mathrm{i} t}=\Delta_{\Psi}^{-\mathrm{i} t} \Delta_{\Psi \mid \Omega}^{\mathrm{i} t} \in \mathcal{A}^{\prime}
\eea
Now, let us return to the problem of state dependence. The exterior right wedge algebra $\mathcal{A}$ is our starting point. Consider two cyclic separating vectors of the algebra, which we will denote by $\Omega$ and $\Psi$. If we can construct two extended  translations using these vectors, we will denote them by $U_{\Omega}(s)$ and $U_{\Psi}(s)$ respectively\footnote{Note that we can always construct an infalling evolution $U_{\Psi}(s)= U U_{\Omega}(s) U^\dagger$ associated to  $\Psi$, where $U$ is the unitary defined by $\Psi = U \Omega$.  These definitions guarantee that $\tilde{\Psi} = U \tilde{\Omega}$. The general properties (like unitarity) of the  $U_{\Psi}(s)$ translation defined this way will be enough for the Connes cocycle proof in this section, to work --  we can be agnostic about the detailed modular translation properties of $U_{\Psi}(s)$. See also the discussion on the frozen vacuum in the final section of this paper to appreciate the naturalness of this construction.}. With the help of these translations, we can construct two different shifted wedge algebras $\mathcal{U}_{\Omega}$ and $\mathcal{U}_{\Psi}$ as in \eqref{shiftedalgebradef}. We can also construct the new cyclic separating vectors $\tilde{\Omega},\tilde{\Psi}$ and modular operators $\tilde{\Delta}_{\Omega}, \tilde{\Delta}_{\Psi}$ associated to these wedges.

Let us look at the crossed products $\mathcal{U}_{\Omega} \rtimes \mathbb{R}_{\Omega}$ and $\mathcal{U}_{\Psi} \rtimes \mathbb{R}_{\Psi}$, where we have adjoined $H_{\Omega,\Psi}/\beta N+X_{\Omega,\Psi}$ to the shifted wedge algebra (Refer appendix \ref{crossedproductrevsec} for more details of the construction). Here $H_{\Omega,\Psi}$ is the modular Hamiltonian of the shifted wedges. Before we proceed, let us define the operator $u_{\Psi \mid \Omega}^{\prime}(P)$ by replacing $t$ with $P = -id/dX$, as in \cite{Witten:2021unn}. This allows us to arrive at our main result:
\begin{center}
\textbf{Theorem 1}: \textit{A unique conjugation relates the two crossed product algebras:}
\bea
\mathcal{U}_{\Psi} \rtimes \mathbb{R}_{\Psi} = \left[U^{\dagger}_{\Psi}(s_0)u_{\Psi\mid\Omega}^{\prime}(P)U_{\Omega}(s_0)\right]\left(\mathcal{U}_{\Omega} \rtimes \mathbb{R}_{\Omega}\right)\left[U^{\dagger}_{\Psi}(s_0)u_{\Psi\mid\Omega}^{\prime}(P)U_{\Omega}(s_0)\right]^{\dagger}\label{crossedproductconjugationeq}
\eea
\end{center}
To prove this statement, let us recall that the crossed product $\mathcal{U}_{\Omega} \rtimes \mathbb{R}_{\Omega}$ is generated by $a_{\Omega}\otimes 1 $ and $\tilde{\Delta}^{-is}_{\Omega}\otimes e^{iX_{\Omega}s}$, where $ a_{\Omega}\in \mathcal{U}_{\Omega}$ and $s \in \mathbb{R}$. Now let us examine the conjugation of the first generator. Dropping the tensor product to simplify notation, we get
\bea
\begin{aligned}
&\left[U^{\dagger}_{\Psi}(s_0)u_{\Psi\mid\Omega}^{\prime}(P)U_{\Omega}(s_0)\right]a_{\Omega}\left[U^\dagger_{\Psi}(s_0)u_{\Psi\mid\Omega}^{\prime}(P)U_{\Omega}(s_0)\right]^{\dagger} \\
&= \left[U^{\dagger}_{\Psi}(s_0)u_{\Psi\mid\Omega}^{\prime}(P)U_{\Omega}(s_0)\right]\left(U_{\Omega}^{\dagger}(s_0)aU_{\Omega}(s_0)\right)\left[U^{\dagger}_{\Omega}(s_0)u_{\Psi\mid\Omega}^{\prime}(P)^{\dagger}U_{\Psi}(s_0)\right]\\
&= U^{\dagger}_{\Psi}(s_0)u_{\Psi\mid\Omega}^{\prime}(P)au_{\Psi\mid\Omega}^{\prime}(P)^{\dagger}U_{\Psi}(s_0)\\
&= U^{\dagger}_{\Psi}(s_0)aU_{\Psi}(s_0)\\
& =a_{\Psi}
\end{aligned}
\eea
where $a \in \mathcal{A}$ and $a_{\Psi} \in \mathcal{U}_{\Psi}$. Here, $u_{\Psi\mid\Omega}^{\prime}(P)$ is an element of the commutant algebra $\mathcal{A}^{\prime}$ that acts on the Hilbert space of small excitations around the TFD state, and we have used $[u_{\Psi\mid\Omega}^{\prime}(P),a]=0$ to go from the third line to the fourth line. Now let us look at the conjugation of the second generator:
\bea
\begin{aligned}
&\left[U^{\dagger}_{\Psi}(s_0)u_{\Psi\mid\Omega}^{\prime}(P)U_{\Omega}(s_0)\right]\left(\tilde{\Delta}^{-is}_{\Omega} e^{iX_{\Omega}s}\right)\left[U_{\Psi}^\dagger(s_0)u_{\Psi\mid\Omega}^{\prime}(P)U_{\Omega}(s_0)\right]^{\dagger} \\
&= \left[U^{\dagger}_{\Psi}(s_0)u_{\Psi\mid\Omega}^{\prime}(P)U_{\Omega}(s_0)\right]\left(U_{\Omega}^{\dagger}(s_0)\Delta^{-is}_{\Omega}U_{\Omega}(s_0)U_{\Omega}^{\dagger}(s_0)e^{iH^{\prime}_Ls/N}U_{\Omega}(s_0)\right)\\
&\hspace{8cm}\left[U^{\dagger}_{\Omega}(s_0)u_{\Psi\mid\Omega}^{\prime}(P)^{\dagger}U_{\Psi}(s_0)\right]\\
&= \left[U^{\dagger}_{\Psi}(s_0)u_{\Psi\mid\Omega}^{\prime}(P)\right]\left(\Delta^{-is}_{\Omega}e^{iH^{\prime}_Ls/N}\right)\left[u_{\Psi\mid\Omega}^{\prime}(P)^{\dagger}U_{\Psi}(s_0)\right]\\
\end{aligned}
\eea
To get to the second line, we have used $\tilde{\Delta}_{\Omega} =U_{\Omega}^{\dagger}(s_0)\Delta_{\Omega}U_{\Omega}(s_0)$ and $e^{iX_{\Omega}s} = U_{\Omega}^{\dagger}(s_0)$ $e^{iH^{\prime}_Ls/N}U_{\Omega}(s_0)$. Now let us look at the following term:
\bea
u_{\Psi\mid\Omega}^{\prime}(P)\left(\Delta^{-is}_{\Omega}e^{iH^{\prime}_Ls/N}\right)u_{\Psi\mid\Omega}^{\prime}(P)^{\dagger}
\eea
In \cite{Witten:2021unn}, this term was shown to be
\bea
\Delta^{-is}_{\Psi}e^{iH^{\prime}_Ls/N}u_{\Psi\mid\Omega}(s)
\eea
This gives us
\bea
\begin{aligned}
&\left[U^{\dagger}_{\Psi}(s_0)u_{\Psi\mid\Omega}^{\prime}(P)U_{\Omega}(s_0)\right]\left(\tilde{\Delta}^{-is}_{\Omega} e^{iX_{\Omega}s}\right)\left[U_{\Psi}^\dagger(s_0)u_{\Psi\mid\Omega}^{\prime}(P)U_{\Omega}(s_0)\right]^{\dagger} \\
&= U^{\dagger}_{\Psi}(s_0)\Delta^{-is}_{\Psi}e^{iH^{\prime}_Ls/N}u_{\Psi\mid\Omega}(s) U_{\Psi}(s_0)\\
&= \left(U^{\dagger}_{\Psi}(s_0)\Delta^{-is}_{\Psi}U_{\Psi}(s_0)U^{\dagger}_{\Psi}(s_0)e^{iH^{\prime}_Ls/N}U_{\Psi}(s_0)\right)\left(U^{\dagger}_{\Psi}(s_0)u_{\Psi\mid\Omega}(s) U_{\Psi}(s_0)\right)\\
&= \left(\tilde{\Delta}^{-is}_{\Psi}e^{iX_{\Psi}s}\right)\left(U^{\dagger}_{\Psi}(s_0)u_{\Psi\mid\Omega}(s) U_{\Psi}(s_0)\right)\\
\end{aligned}
\eea
Since $u_{\Psi\mid\Omega}(s) \in \mathcal{A}$, the final expression is a product of two elements of the crossed product $\mathcal{U}_{\Psi} \rtimes \mathbb{R}_{\Psi}$. Therefore, the resulting operator is also an element of the crossed product, and this completes our proof of \eqref{crossedproductconjugationeq}. In the above steps we have used the fact that the relevant calculations in Section \ref{mech} go through even when $X (= X_{\Omega})$ is replaced by $X_{\Psi}$. The conjugations with $U(s_0) (=U_{\Omega}(s_0))$ can all be suitably replaced by conjugations with $U_{\Psi}(s_0)$.

\subsection{Weak vs Strong Forms of State-Independence}\label{WvsS}

Let us take a moment to clarify what we mean by state-independence in these discussions. 

There is a sense in which one might think that state-independence in the interior is already present at the level of the Type III algebras. The reason is that given an initial right wedge algebra and two different cyclic separating vectors $\Omega$ and $\Psi$, we can construct translations $U_\Omega(s)$ and $U_\Psi(s)$ corresponding to them. This means that there is a unique isomorphism $U^\dagger_\Omega(s) U_\Psi(s)$ between the two resulting interior algebras. Since there is a unique isomorphism, in some sense, these are the same algebras. We will call this weak state-independence.

Weak state-dependence does not qualify as meaningful state-independence. The reason is that without discussing an appropriate form of dynamics, it is not very meaningful to compare these algebras. It is a bit like comparing Hilbert spaces without discussing their Hamiltonians. They are all isomorphic if their dimensions are the same, but not (necessarily) in a way that respects dynamics. 

However, if we simply adjoin a modular Hamiltonian associated to a cyclic separating vector to a Type III$_1$ algebra, the resulting algebra is not independent of the state. This also means that such an algebra after modular translation will again have a dependence on the choice of state. This is what makes the crossed product interesting, and this is why we needed the Connes cocycle to prove that the modular translated algebra is indeed state-independent. We can think of this as a strong form of state-independence and this is the useful notion of state-independence that is relevant to this paper. As discussed in Sections \ref{intro-sub} and \ref{summary}, this makes interior reconstruction precisely as state-independent as exterior reconstruction. 

\subsection{Subalgebras and Emergent Symmetries}\label{subalgebra}

Consider two subalgebras $\mathcal{N}, \tilde{\mathcal{N}} \subset \mathcal{A}_R$. Operationally, one can obtain these subalgebras by restricting the operators in $\mathcal{A}_R$ to different subregions $\mathcal{N},\tilde{\mathcal{N}}$ in the exterior right wedge of the spacetime. Given these subalgebras, we can construct two \textit{distinct} extended modular translations $U(s)$ and $\tilde{U}(s)$ satisfying the initial conditions\footnote{Note that we are working with the same cyclic separating vector for all modular translations in this subsection.}
\bea
\mathcal{N}=U^{\dagger}(-1)\mathcal{A}_R U(-1) \quad \quad \quad \tilde{\mathcal{N}}=\tilde{U}^{\dagger}(-1)\mathcal{A}_R \tilde{U}(-1).
\eea
Using the two extended modular translations, we can define two different shifted wedge algebras, which we will label by $\mathcal{U}$ and $\tilde{\mathcal{U}}$. Let us assume that these modular translations map two different points $x_R \in \mathcal{N}$ and $\tilde{x}_R \in \tilde{\mathcal{N}}$ to the same point in the interior, which we will label by $x$. We have
\bea
\phi(x) = U^{\dagger}(s_0)\phi(x_R) U(s_0) =  \tilde{U}^{\dagger}(s_0)\phi(\tilde{x}_R) \tilde{U}(s_0)
\eea
Inverting the unitary operators, we get
\bea
\phi(x_R) =\left(U(s_0)\tilde{U}^{\dagger}(s_0)\right)\phi(\tilde{x}_R) \left(U(s_0)\tilde{U}^{\dagger}(s_0)\right)^{\dagger}.\label{subalgebraunitaryconjugation}
\eea
Therefore, the operators acting at different points in the exterior algebra can be mapped to one another using a unitary operator $U(s_0)\tilde{U}^{\dagger}(s_0)$. 

It is natural to suspect that such unitaries correspond to symmetries of the bulk spacetime. It will be interesting to see if one can identify useful notions of (emergent) symmetries beyond the isometries of the background, by such a procedure in holographic theories. 

\section{Thermality, Typicality, Firewalls and $N$}\label{more}

The discussion above was in the context of the two-sided black hole, which uses the thermofield double as the cyclic separating vector. This is slightly different from the setting of \cite{Papadodimas:2013jku} who instead works with a heavy microcanonical typical pure state of a single CFT, $|\Psi_0\rangle$, that is close to thermality \cite{Papadodimas:2013jku}. The analogue of the right wedge algebra here is straightforward to define: it is the algebra of single trace operators $\mathcal{A}_R$ in the large-$N$ limit\footnote{According to the recent results of \cite{Leutheusser:2021frk, Witten:2021unn} it is in the large-$N$ limit that a smooth black hole interior along with a commutant structure emerges. The story is quite remarkable, so we provide a small review of the relevant facts. Our discussion of \cite{Papadodimas:2013jku} will be in that language in what follows.}. We assume that $|\Psi_0\rangle$ is cyclic and separating\footnote{See section V E of \cite{Leutheusser:2022bgi} for some related recent discussions invoking time-bands.}, and that the algebra $\mathcal{A}_R$ acts reducibly on $|\Psi_0\rangle$. We will denote the commutant by $\mathcal{A}_L$ to emphasize the parallel with the thermofield double. These assumptions are natural in the large-$N$ limit because of the emergent Type III$_1$ structure \cite{Leutheusser:2021frk} of the single trace sector. In this setting one can use Tomita-Takesaki theory to construct mirror operators in $\mathcal{A}_L$ for operators in $\mathcal{A}_R$, by conjugating the right wedge operators using the modular conjugation $J$. If one further {\em assumes} that bulk Kruskal evolutions are well-defined, one can then use these operators to construct bulk local operators in the black hole interior. See eg., eqn (2.4) of \cite{Papadodimas:2013jku} where the time coordinate is an interior bulk time.

The crucial realization of \cite{Leutheusser:2021frk, Witten:2021unn} is that the emergence of the type III$_1$ algebra in the large-$N$ limit is intimately tied to the emergence of the interior in that limit. At finite-$N$, the entire CFT algebra is generated by the single trace sector and therefore it is Type I. But at large-$N$ the single trace sector becomes Type III$_1$, and Type III algebras have non-trivial commutants. The commutant is what gets interpreted as the operator algebra responsible for the interior of the black hole. In the Type III$_1$ setting, the interior time can be constructed via modular translations \cite{Leutheusser:2021frk} and we expect that bulk Kruskal-like times can be justified instead of being postulated. We will not try to write down here the modular translation in a general Kruskal-AdS geometry that reproduces a conventional interior time -- this is likely too complicated to be analytically tractable. But we will write down a pointwise, fully analytic modular translation that takes us inside the BTZ horizon in the next section.

A related fact is that while construction of TFD-like states is possible at finite-$N$ in doubled CFTs, their interpretation as smooth Kruskal spacetimes is straightforward only in the large-$N$ limit -- see eg. the discussion regarding non-analyticity at large-$N$ in \cite{Leutheusser:2021qhd}. In other words, \cite{Leutheusser:2021frk, Leutheusser:2021qhd} argue that the smooth horizon of a thermal state is the result of a large-$N$ phase transition and the thermodynamic limit (see also \cite{WittenQFT}). The Type III modular translations of \cite{Leutheusser:2021frk} justify the interior time in the large-$N$ limit and the construction of \cite{Witten:2021unn} allows us to move perturbatively away from large-$N$. 

The statement that the exterior algebra is Type I at finite-$N$ has found legitimacy in recent papers (see eg., \cite{Witten:2021unn}). This is a (more intrinsically quantum) version of the older brick-wall/stretched horizon \cite{tHooftBrickWall, Larus} or fuzzball \cite{LuninMathur} idea, interpreted as the statement that the UV complete {\em bulk} description of a black hole microstate does not have an interior (ie., a non-trivial commutant). In particular, these recent developments are consistent with the possibility that typical states in the CFT have firewalls \cite{MarolfPolchinski1}\footnote{But let us also note \cite{Corley} which may provide a loophole.}. It has been suggested \cite{CPW} that the emergence of the Type III$_1$ structure is best understood by viewing it as a statement about complexity -- if we only consider operators of complexity below some threshold while first sending $N$ to infinity, those are the operators that form the Type III$_1$ algebra (once you allow the threshold to go to infinity afterwards). This observation is compatible with the claims \cite{Susskind, Akers} that only horizons with increasing complexity can be smooth. The idea here is that we should only consider products of upto a maximum number $k$ of single trace operators when sending $N$ to infinity. This is a parametric bound on the complexity of operators. After $N$ has gone to infinity, one can send $k$ to infinity. If the large-$N$ limit is interpreted in this way, it is straightforward to see that complexity never saturates in such a set up. This is because complexity saturates at $e^{O(S_{BH})}$, which is infinite in the large-$N$ limit. This explains the smoothness of the horizon of the thermofield double. Since typical states saturate complexity, it also means that these large-$N$ eternal black holes where complexity never saturates should not be viewed as typical. At finite $N$, even though the single trace algebra is not really Type III$_1$, it is reasonable to expect that an approximation that trades off between $k$ and $N$ should be able to see an approximately smooth horizon when the black hole is (thermal, but) not typical. Note that this also means that the ``eternal'' black hole geometry is only valid for a finite amount of time at finite-$N$. We will elaborate on some of these comments in an upcoming paper \cite{KM}.


Another (more minor) distinction between \cite{Papadodimas:2013jku} and \cite{Leutheusser:2021frk} is that the former works more closely with the representation of the algebra than the algebra itself, by choosing a specific cyclic separating vector and working with mirror operators constructed from them. Equivalent choices are implicit in the construction of \cite{Leutheusser:2021frk} as well, because the modular translations depend on both the right and left wedge algebras. As we noted earlier, the construction of \cite{Papadodimas:2013jku} works in practice by solving bulk wave equations, but these distinctions are less important from our perspective. The state dependence of the interior operator construction is shared by both approaches. 


\section{Extended Modular Translations in BTZ Black Holes}
\label{BTZ}

In this section we will explicitly construct certain extended translation $U(s)$ that takes operators across the horizon of a BTZ black hole \cite{Banados:1992wn,Banados:1992gq}. In comparison to \cite{Leutheusser:2021frk}, where the modular translation was only pointwise in the near-horizon limit, our construction is analytical and applies point-wise (not non-local) throughout the entire BTZ spacetime. The price we pay is that it is a somewhat unintuitive modular translation when viewed from the BTZ-Kruskal coordinates, because its origins are to be traced to the covering space of BTZ and a connection with AdS-Rindler (whose modular translations we will exploit)\footnote{\label{generalized-mod}In other words, these are descendants of the covering space modular translations in the quotient BTZ. In the BTZ geometry, they are more accurately called ``generalized'' modular translations -- but we will suppress this neologism. The key feature we will be interested in is that these are unitaries that can evolve into the black hole interior.}.

We will construct the BTZ black hole by quotienting the Poincar\`{e} patch of AdS$_3$. Then, we will show that the modular inclusions constructed in \cite{Leutheusser:2021frk} can evolve operators across the BTZ horizon.
\subsection{BTZ as a quotient of the Poincar\'{e} Patch of AdS}
In the Schwarzschild coordinates, the BTZ metric takes the form:
\bea
d s^{2}=-\frac{r^{2}-r_{S}^{2}}{R^{2}} dt^{2}+\frac{R^{2}}{r^{2}-r_{S}^{2}} d r^{2}+r^{2} d \phi^{2} \quad \quad(\phi \sim \phi+ 2 \pi, r>0) \label{schwarzschildmetric}
\eea
The horizon of the black hole is at $r=r_S$ while the singularity is at $r=0$. We will work with the following rescaled coordinates \cite{delaFuente:2013nba}
\bea
\frac{R}{r_S}r \to r, \quad \quad \frac{r_S}{R}t \to t, \quad \quad \phi \to r_S \phi,
\eea
We will be working with $R=1$ units in the rest of this section, and therefore $r_S=r_S/R$ is dimensionless and the horizon is at $r=1$. Note however that the physical horizon radius is still arbitrary and captured by the periodicity of $\phi$ which is $2 \pi r_s$.  We will sometimes set $r_S=1$ in making plots, accepting some loss of generality. We can extend the spacetime by using the Kruskal coordinates. The extended spacetime, which we will refer to as BTZ$_{\text{Kruskal}}$, has the metric
\bea
d s^{2}=-\frac{4 d u d v}{(1+u v)^{2}}+\left(\frac{1-u v}{1+u v}\right)^{2} d \phi^{2} \quad \quad (|uv|<1,\phi \sim \phi+ 2\pi r_s) \label{Kruskalmetric}
\eea
Note again the range of $\phi$. The event horizon is located at $u,v=0$, while the singularity is at $uv =1$. The asymptotic boundaries can be found at $uv=-1$.

BTZ geometry can also be understood as a quotient of a covering AdS$_3$ space. The surface $uv=1$ is a quotient singularity and not a true curvature singularity. The structure of the quotient is particularly transparent in the Poincar\'{e} patch of AdS \cite{delaFuente:2013nba,Maldacena:1998bw,Carlip:1995qv,Danielsson:1998wt}:
\bea
\text{BTZ}_{\text{Poincar\'{e}}} = \text{AdS}_{\text{Poincar\'{e}}}/\Gamma, \label{quotientglobal}
\eea
Here $\Gamma$ is a quotient discrete isometry group of AdS. The resulting spacetime includes BTZ$_{\text{Kruskal}}$ as well as two \textit{whisker} regions beyond the singularity. To discuss the quotient, let us start with $\text{AdS}_{\text{Poincar\'{e}}}$
\bea
d s^{2}=\frac{1}{z^2}\left(-dx^{+}dx^{-}+dz^2\right) \quad \quad(z>0)
\eea
and then identify points by the following discrete rescaling
\bea
x^{\pm} \equiv e^{r_S}x^{\pm} \quad \quad z \equiv e^{r_S} z. \label{poincarequotienteq}
\eea
We will identify $\text{BTZ}_{\text{Poincar\'{e}}}$ as the fundamental domain $\mathscr{F}$ of this operation, which we will choose to be the region within the coordinate ranges
\bea
x^{+}\in \left(-e^{r_S},-1\right] \cup \{0\} \cup \left[1,e^{r_S}\right). \label{fundamentaldomainrangeeq}
\eea
It is easy to see why $\mathscr{F}$ is a fundamental region. Consider some point $x_0 =(x^{+}_0,x^{-}_0,z_0) \notin \mathscr{F}$. Under the quotienting \eqref{poincarequotienteq}, $x_0$ will get identified with another point $x_1 =e^{-r_S} (x^{+}_0,x^{-}_0,z_0)$, which will in turn be identified with $x_2 =e^{-2r_S} (x^{+}_0,x^{-}_0,z_0)$ and so on. Eventually, we will reach a value of $k$ for which the $x^{+}$ coordinate of the point $x_k$ will belong to the range indicated in \eqref{fundamentaldomainrangeeq}. Therefore, for any given point $x_0$ in the Poincar`{e} patch, we can always find a point $x_k \in \mathscr{F}$ which belongs to the orbit of $x_0$. Moreover, no two points in $\mathscr{F}$ can be mapped to one another under the rescaling \eqref{poincarequotienteq}. This shows that $\mathscr{F}$ is a fundamental domain.

Our approach in the following will be to use modular translations in the covering space to learn things about the quotient BTZ. For this, it will be very useful to note that the Poincar\`{e} coordinates are related to the Schwarzschild coordinates as follows:
\bea \label{BTZ2Poincare}
x^{ \pm}=\left\{\begin{array}{c}
\pm \sqrt{1-\frac{1}{r^2}} e^{ \pm \xi^{ \pm}}, \quad \text { if } r>1 \\
\phantom{\pm}\sqrt{\frac{1}{r^2}-1} e^{ \pm \xi^{ \pm}}, \quad \text { if } r<1 \\
\end{array}\right. ; \quad \quad z = \frac{e^{\phi}}{r}
\eea
where $\xi^{\pm} = t \pm \phi$.  In terms of the Schwarzschild coordinates, the quotienting described above becomes the periodicity of the $\phi$-direction. A crucial point that follows from this, is that if we do {\em not} impose the periodicity in the $\phi$-direction, the (exterior) Schwarzschild form of BTZ is {\em identical} to the AdS-Rindler region(s) that constitute AdS-Poincare. This means that except for the discreteness vs. continuousness of the angular momentum label $k$, the mode functions of both these geometries are the same. See the discussion at the beginning of Section VIII B in \cite{Leutheusser:2021frk}. It follows that we can define a natural class of (infalling time) translated mode functions for BTZ, by starting with the modular translated mode functions of AdS-Rindler constructed in \cite{Leutheusser:2021frk}, but with the understanding that the $k$-label needs to be discrete\footnote{See the general discussion (see eg., Section VI) of \cite{Leutheusser:2021frk} to see the general structure of modular translations on generalized free fields.}.  This is what we will exploit in the next subsection.

An important consequence of the quotient construction is that $\text{BTZ}_{\text{Poincar\'{e}}}$ contains BTZ$_{\text{Kruskal}}$. We can make this connection explicit by relating the Poincaré and Kruskal coordinates:
\bea
x^{+}=\frac{2 e^{\phi} v}{1-u v} \quad \quad x^{-}=\frac{2 e^{\phi} u}{1-u v}\quad \quad z=\frac{1+u v}{1-u v} e^{\phi}.\label{poincaretokruskaleq}
\eea
The BTZ$_{\text{Kruskal}}$ boundaries, singularity, and horizons get mapped to the following points \cite{delaFuente:2013nba}:
\begin{figure}
\centering
\begin{subfigure}[b]{0.45\textwidth}
\centering
\includegraphics[width=\textwidth]{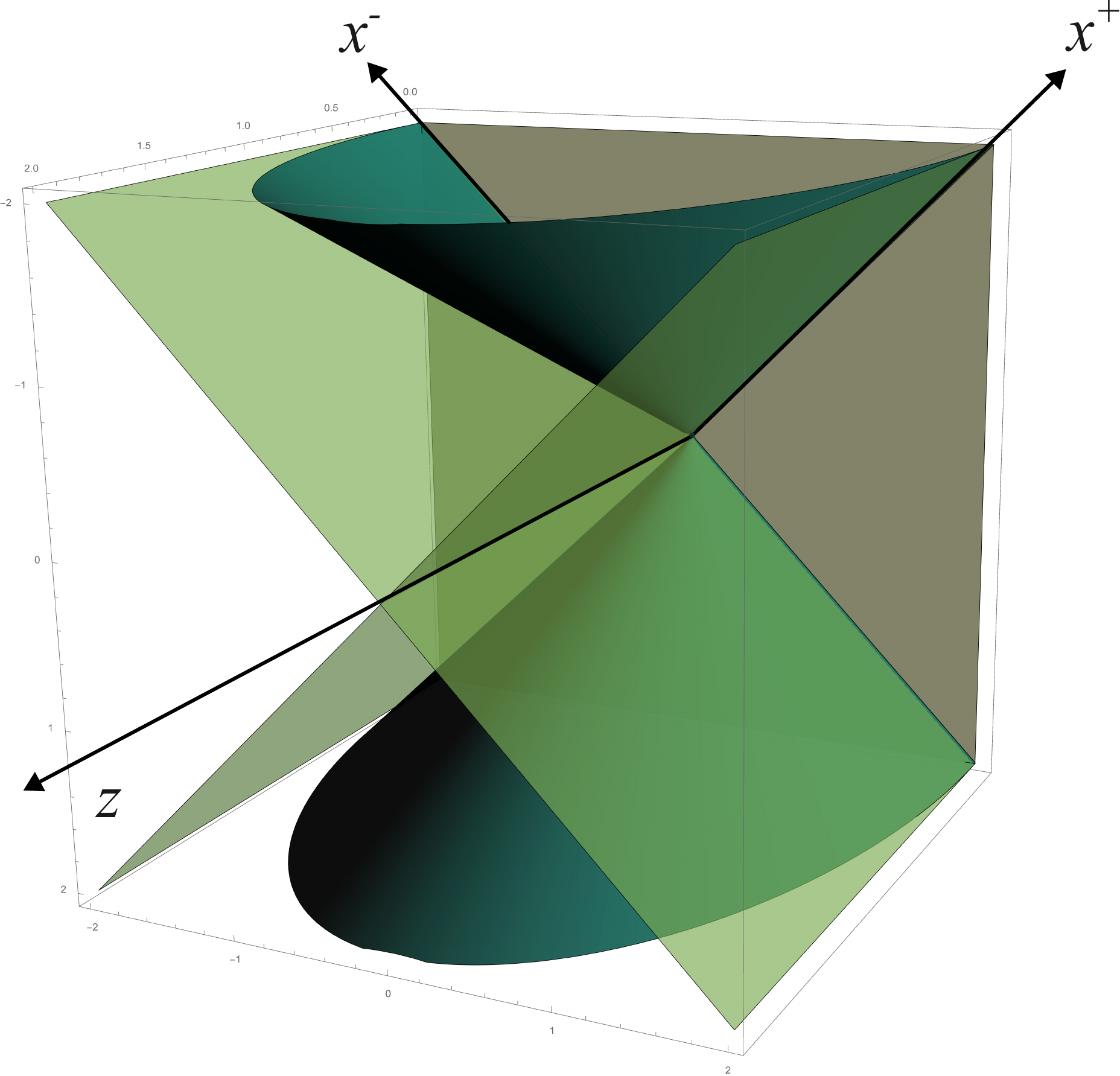}
\caption{}
\label{}
\end{subfigure}
\hfill
\begin{subfigure}[b]{0.45\textwidth}
\centering
\includegraphics[width=\textwidth]{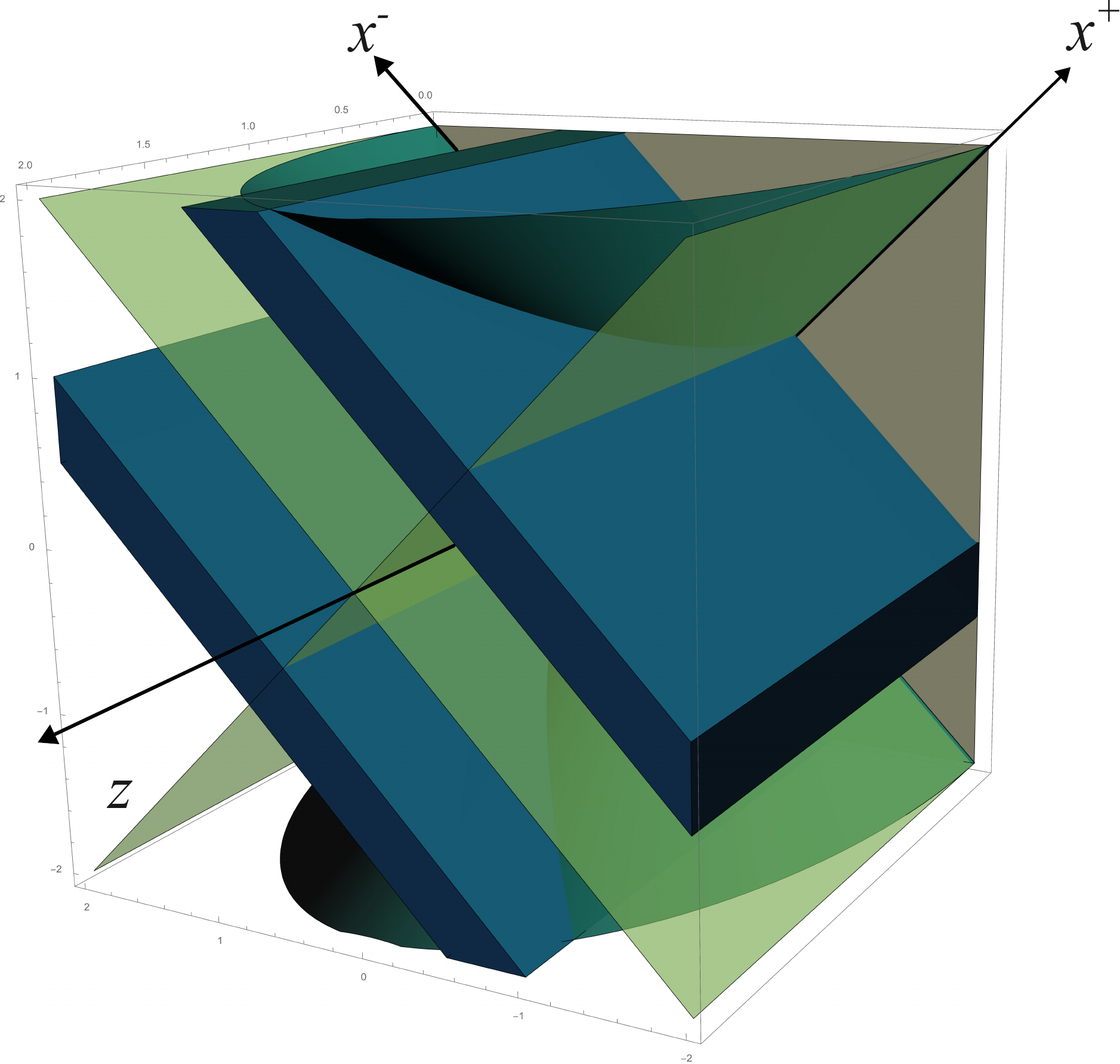}
\caption{}
\label{}
\end{subfigure}
\caption{(a) The plot shows AdS$_{\text{Poincar\'{e}}}$. The green null sheets become the horizons after quotienting. The singularity emerges from the cones, while the boundary comes from the $z = 0$ plane. After identifying points, the region of spacetime within the cones becomes what we refer to as the ``whisker'' regions. (b) The blue boxes, along with the $x^{+}=0$ surface, correspond to the fundamental domain $\mathscr{F}$.}\label{BTZpoincarefig}
\end{figure}

\bea
\begin{aligned}
\text{boundary:}& \quad z=0 \\
\text{horizon:}& \quad x^{\pm}=0\\
\text{singularity:}& \quad z^{2}-x^{+} x^{-}=0.
\end{aligned}
\eea
We can identify the region $z^{2}-x^{+} x^{-}>0$ as the BTZ$_{\text{Kruskal}}$ spacetime, excluding the whisker regions (we will not be concerned with the latter). Plotting these surfaces, we get figure \ref{BTZpoincarefig}.

We can also partition the spacetime into four regions:
\bea
\begin{aligned}
\text{Right} (\mathcal{R}):\quad &x^{+}\in \left[1,e^{r_S}\right), \phantom{--}\quad x^{-}<0\\
\text{Future} (\mathcal{F}): \quad &x^{+}\in \left[1,e^{r_S}\right), \phantom{--}\quad x^{-}>0\\
\text{Left} (\mathcal{L}): \quad &x^{+}\in \left(-e^{r_S},-1\right], \quad x^{-}>0\\
\text{Past} (\mathcal{P}): \quad &x^{+}\in \left(-e^{r_S},-1\right], \quad x^{-}<0.
\end{aligned}\label{wedgeseq}
\eea
We can see that the regions $\mathcal{R},\mathcal{L}$ lie in the exterior of the horizon while the regions $\mathcal{F},\mathcal{P}$ are in the interior of the black hole. A key point we will exploit in our construction of the modular translation is that we can cross the horizon by shifting $x^-$ while staying within this fundamental region.

Before we conclude this subsection, let us make a couple of observations:
\begin{itemize}
\item The Poincar\`{e} BTZ contains the Kruskal BTZ:
\bea
\text{BTZ}_{\text{Poincar\'{e}}} \supset \text{BTZ}_{\text{Kruskal}}.\label{kruskalpoincaresubseteq}
\eea
Therefore, we can always use the Poincar\`{e} coordinates to cover the Kruskal spacetime.
\item Since the quotienting \eqref{poincarequotienteq} involves only discrete rescaling, the BTZ horizon emerges from the AdS-Rindler horizon. This fact will turn out to be quite useful in the next subsection.
\end{itemize}

\subsection{Half-sided modular translations in Poincar\'{e} BTZ}
\label{modularsection}
Now let us construct an extended half-sided modular translation $U(s)$ in $\text{BTZ}_{\text{Poincar\'{e}}}$. The action of $U(s)$ is completely specified up to a phase $e^{i\gamma_k}$ (See appendix \ref{modulartransappendix} for more details). We will formally choose the same phase used in the AdS-Rindler calculation in \cite{Leutheusser:2021frk}. Even though the resulting operator $U(s)$ has the same form as the one in \cite{Leutheusser:2021frk}, the $k$-index now leads to a discrete sum. The new operator $U(s)$ has a natural interpretation as a bulk evolution for two essential reasons. Firstly, $\text{BTZ}_{\text{Poincar\'{e}}}$ is the fundamental region $\mathscr{F}$ of the Poincar\`{e} patch under the quotienting that leads to the quantization of $k$. As a result, the $U(s)$ has a natural induced action on the operators in $\text{BTZ}_{\text{Poincar\'{e}}}$. Secondly, we will see that the operator $U(s)$ generates translations of the form $x^{-} \to x^{-}+s$, and the shifted points remain within the same fundamental domain. Therefore, we can view $U(s)$ as a generalized (extended) half-sided modular translation in the Poincar\'{e} BTZ spacetime. As mentioned in footnote \ref{generalized-mod}, our primary goal here is to come up with  a unitary that leads to bulk evolutions that takes us inside the horizon.

In this section, we will see that these translations act \textit{point-wise} and that they push operators across the horizon of the black hole.
Consider a bulk scalar field $\phi$ with scaling dimension $\Delta$. If we restrict the scalar field to the $\mathcal{R}$ region of the BTZ black hole, then we can expand it in terms of the normalizable modes as follows\footnote{We will follow the notations and conventions in section IX of \cite{Leutheusser:2021frk}. Since our aim in this section is not to redo the calculations in \cite{Leutheusser:2021frk}, we will only quote the relevant results here. Interested readers can refer to \cite{Leutheusser:2021frk} for more details. We found it convenient to do some of the calculations in \cite{Leutheusser:2021frk} using Pochhammer symbol manipulations than Appel functions.}:
\bea
\phi_{R}(X)=\sum_{k} v_{k}^{(R)}(X) a_{k}^{(R)},\label{fieldexpansioneq}
\eea
Now let us look at the evolution of the scalar field under $U(s)$:
\bea
\Phi(X ; s)=U(s)^{\dagger} \phi_{R}(X) U(s)\label{eveolutioneq}
\eea
The evolution is completely fixed up to a phase factor $e^{i\gamma_k}$ and we will adopt the same form as in section IX of \cite{Leutheusser:2021frk}. As we discussed earlier, a crucial fact here is that the mode functions in $\text{BTZ}_{\text{Poincar\'{e}}}$ and AdS-Rindler have the same form owing to the quotienting structure, except for the fact that the angular quantum number is discrete in BTZ. Therefore we can carry over the calculations in section IX of \cite{Leutheusser:2021frk} with the angular quantum number label understood as discrete. This is what we will mean by $U(s)$ in the context of the BTZ geometry. We can then interpret these answers in BTZ coordinates $(t,r, \phi)$\footnote{The translation between BTZ and AdS-Rindler is given in \eqref{BTZ2Poincare}.}. In particular, when
\bea
|s|<s_1 \equiv e^{-t+\phi}\sqrt{1-\frac{1}{r^{2}}},\label{eqs0}
\eea 
we find that
\bea
\Phi(X;s)= \phi_R(X_s), \quad s <s_1
\eea
\begin{figure}
\centering
\begin{subfigure}[b]{0.45\textwidth}
\centering
\includegraphics[width=\textwidth]{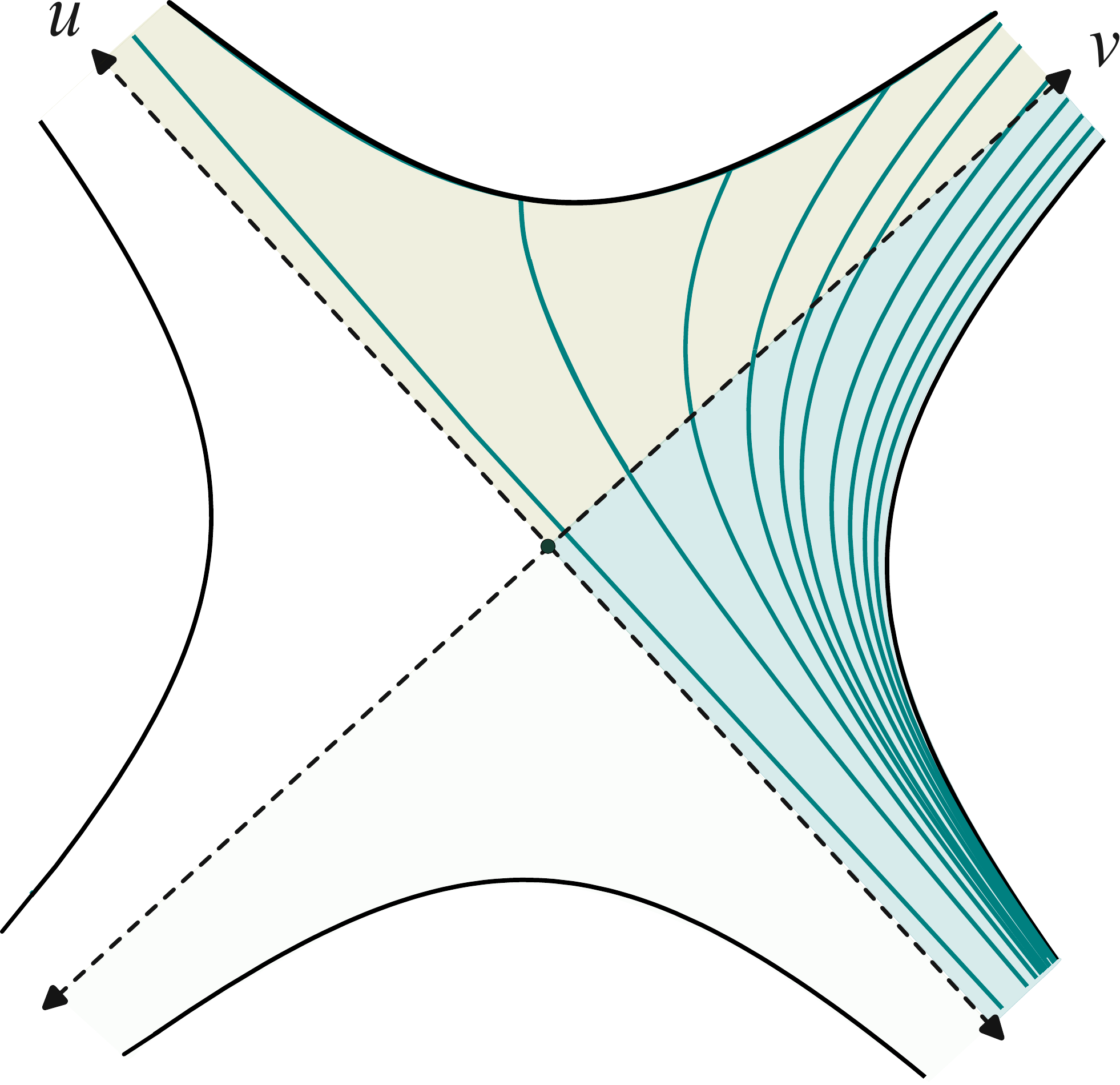}
\caption{}
\label{flowdiagramKruskal}
\end{subfigure}
\hfill
\begin{subfigure}[b]{0.45\textwidth}
\centering
\includegraphics[width=\textwidth]{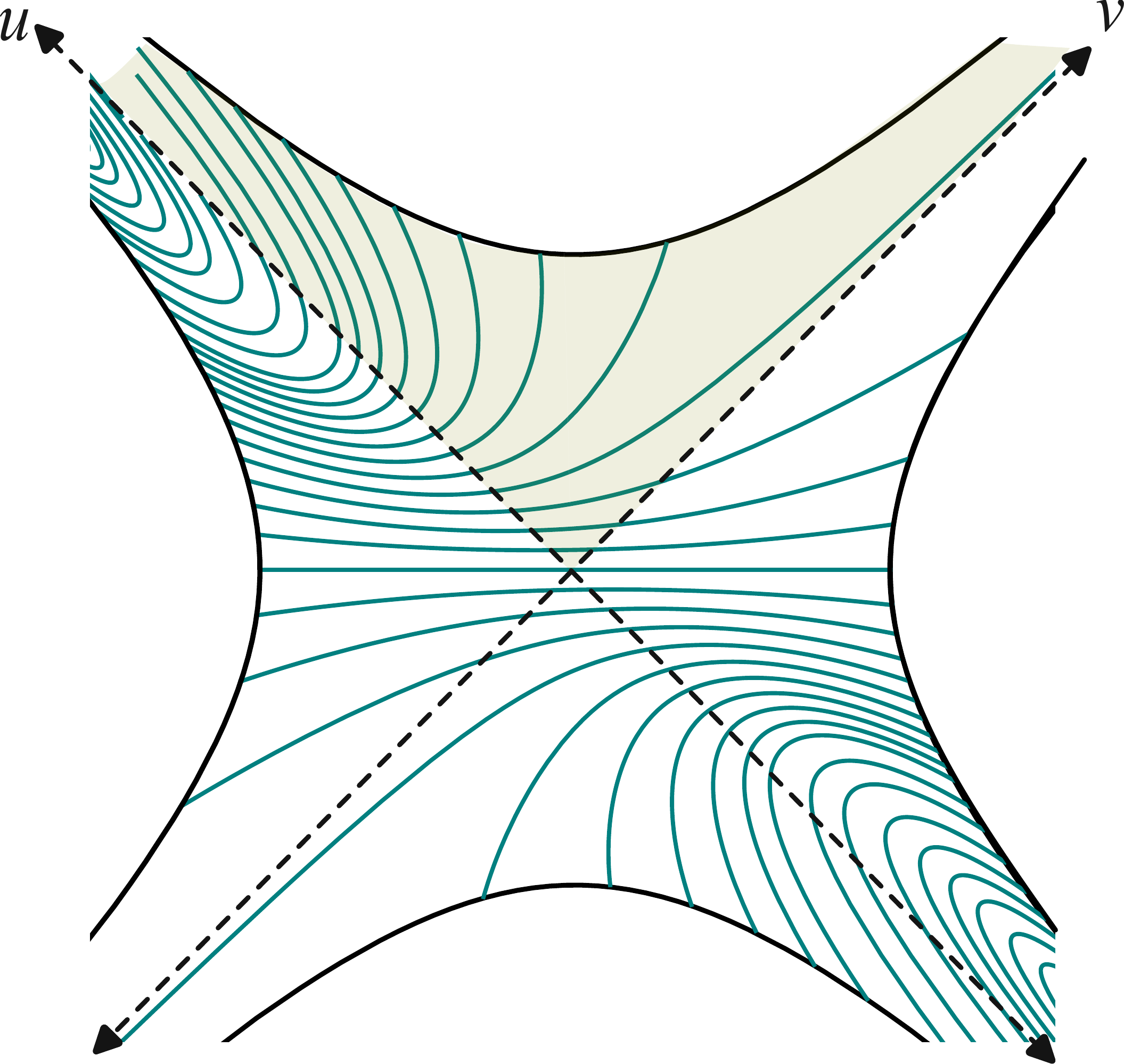}
\caption{}
\label{s-flowdiagramKruskal}
\end{subfigure}
\caption{(a) The plot shows the flow of operators from the exterior right wedge to the interior of the black hole as they evolve under $U(s)$. (b) The plot gives the evolution of a constant $t=0$ slice under the action of $U(s)$. We have chosen the initial value of $\phi$ to be zero here.}
\end{figure}
where $X_s=(t_s,r_s,\phi_s)$ are given by\footnote{The translated radial coordinate $r_s$ should not be confused with the horizon radius $r_S$. We will set $r_S=1$ when making plots.}
\bea
\begin{aligned}
r_s = r\sqrt{1-a_s} \ \ & \ \quad a_s \equiv se^{t-\phi}\sqrt{1-\frac{1}{r^{2}}}\\
e^{t_s} = e^t\sqrt{\frac{r^2-1}{r^2(1-a_s)-1}} \ \ & \ \quad e^{\phi_s} = e^\phi\sqrt{1-a_s}
\end{aligned}\label{globaltopoincareshifteq1}
\eea
These are simply the modular translations discussed in \cite{Leutheusser:2021frk} in Poincar\'{e} coordinates, which take the simple form 
\bea
(x^+,x^-,z) \to (x^+,x^-+s,z)\label{revolpoin}
\eea
translated to the BTZ Schwarschild coordinates. 
As $s<s_1\equiv-x^{-}$, $X_s \in \mathcal{R}$. Therefore, the evolved operator is still in the right exterior wedge.

When $s \in (s_1,s_2)$, where
\bea
s_2  \equiv e^{-t+\phi}\frac{1}{\sqrt{1-\frac{1}{r^{2}}}},\label{eqs1}
\eea
we get
\bea
\Phi(X;s)= \phi_F(X_s), \quad s \in (s_1,s_2)
\eea
where $\phi_F$ is the bulk local field in the interior of the black hole (Refer \cite{Leutheusser:2021frk} for more details). Here $X_s=(t_s,r_s,\phi_s)$ is given by
\bea
\begin{aligned}
r_s = r&\sqrt{1-a_s} \\
e^{t_s} = e^t\sqrt{\frac{r^2-1}{r^2(a_s-1)-1}} \ \ & \ \ e^{\phi_s} = e^\phi\sqrt{1-a_s}
\end{aligned}\label{globaltopoincareshifteq2}
\eea
In the Poincar\'{e} coordinates, these transformations can be written as $X_s = (x^+,x^{-}+s,z)$. Moreover, we can see that
\bea
s>s_1 = -x^{-} \quad \implies \quad X_s \in \mathcal{F}
\eea
When $s=s_2$, the operator hits the singularity. Therefore, the evolution under $U(s)$ has taken the bulk local operator across the horizon. Moreover, $s_1$ turns out to be the horizon crossing time.

\begin{figure}
\centering
\includegraphics[width=1.03\linewidth]{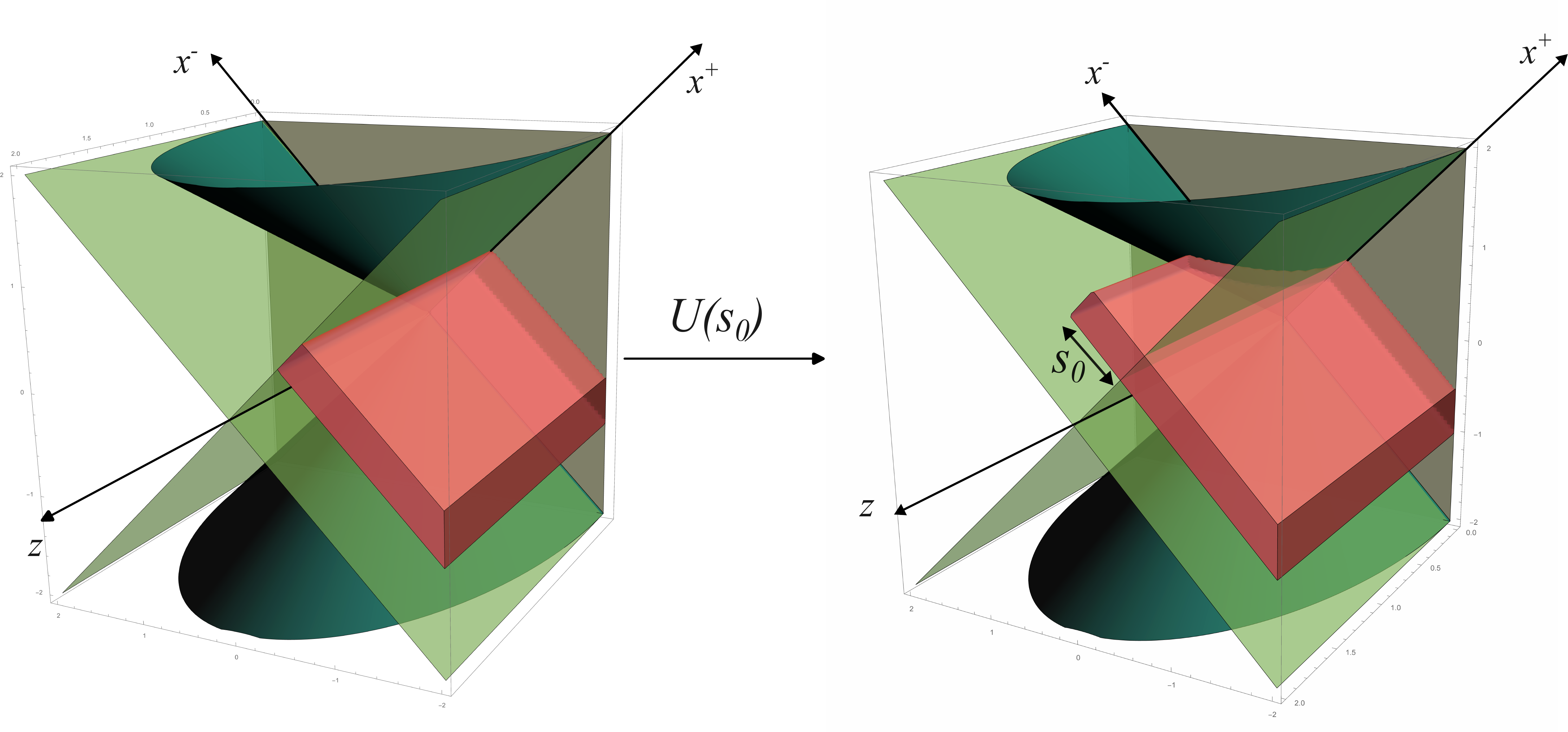}
\caption{Consider the right exterior wedge of the BTZ black hole defined in \eqref{wedgeseq}. The red area in the left figure indicates this region. Under the action of $U(s_0)$, for some $s_0>0$, the wedge gets shifted into the interior. This gives us the figure on the right. We can see that there are some points in the exterior which gets pushed into the region behind the singularity. These points, after the translation, are present only in Poincar\`{e} BTZ, not Kruskal BTZ and should be thought of as a part of the whisker region \cite{delaFuente:2013nba}.}
\label{shiftedwedgefig}
\end{figure}

Now, let us examine the evolution in Kruskal coordinates. Using \eqref{poincaretokruskaleq}, we find that
\bea
\begin{aligned}
u_s &= \frac{1+uv}{2v}-\frac{1-uv}{2v}\sqrt{1-\frac{2sve^{-\phi}}{1-uv}} \\
v_s &= \frac{e^{\phi}}{s + 2 e^{\phi} u - s u v}\left(1+uv-(1-uv)\sqrt{1-\frac{2sve^{-\phi}}{1-uv}}\right)\\
e^{\phi_s} &= e^\phi\sqrt{1-\frac{2sve^{-\phi}}{1-uv}}.
\end{aligned}
\eea
Plotting these curves on the Kruskal diagram, we get the flow in figure \ref{flowdiagramKruskal}, and we can see that the modular translations take us into the interior of the black hole. Note that the coordinate transformation \eqref{poincaretokruskaleq} is non-linear. Inverting the equations, we obtain
\bea
u = \frac{z-\sqrt{z^2-x^{+}x^{-}}}{x^{+}}, \quad v = \frac{z-\sqrt{z^2-x^{+}x^{-}}}{x^{-}}, \quad \phi = \log{\left(\sqrt{z^2-x^{+}x^{-}}\right)}
\eea
The fundamental domain in the Poincar\`{e} coordinates does not get mapped to a single fundamental domain in the Kruskal coordinates. So we have to identify $\phi \sim \phi+ 2\pi$ (we have set $r_S=1$) to make sense of these translations in the Kruskal coordinates. Now, let us look at the evolution of the exterior wedge under the evolution of $U(s)$. For some $s_0>0$, the boundary of the exterior wedge gets pushed into the interior. This gives us figure \ref{shiftedwedgefig}.


Now let us very briefly look at the evolution of the boundary operators under the action of $U(s)$. Consider an operator $\mathcal{O}_R(x)$ dual to the bulk operator $\phi_R(X)$. We will use $x$ to denote the coordinates of the boundary theory. We can define the boundary operator using the extrapolate dictionary in the exterior region. By stripping off a factor of $r^{-\Delta}$ and taking the $r \to \infty$ limit, we get
\bea
\mathcal{O}_R(x)=\sum_k u_k^{(R)}(x) a_k^{(R)}, \quad u_k^{(R)}(x)=N_k e^{-i \omega t+i q \phi}.
\eea
The evolution under $U(s)$ is given by
\bea
\begin{gathered}
\mathcal{O}_R(x ; s)=U(s)^{\dagger} \mathcal{O}_R(x) U(s)=\sum_{k^{\prime}} u_{k^{\prime}}^{(R \beta)}(x ; s) a_{k^{\prime}}^{(\beta)}
\end{gathered}
\eea
where $u_{k^{\prime}}^{(R \beta)}(x ; s)$ are the evolved mode functions. The action of $U(s_0)$ on the boundary operator can be explicitly worked out, as in section VII of \cite{Leutheusser:2021frk} which presents the result for boundary Rindler. Note that there is a crucial difference here. We are looking at the operators living on the $r\to \infty$ boundary, while the boundary operators in section VII of \cite{Leutheusser:2021frk} live on the Poincaré boundary. This results in an additional factor of $e^{\Delta \phi}$ in the boundary mode functions $u_k^{(R)}(x)$ (See appendix E of \cite{Leutheusser:2021frk} for more details).  

It will be interesting to develop this systematically for boundary representatives of interior evolutions as well, but we will not do so here.

\subsection{The Singularity}\label{singularity}

As we mentioned earlier, the results of this section may be of some technical utility in that they provide a point-wise, analytically tractable modular translation through a black hole horizon. One immediate use of such a modular translation is that we can access the singularity using it. Note that constructions that rely on the near horizon geometry cannot be used reliably to go deep into the black hole. 

We plot the constant $s$-slices associated to our modular flows in Figure \ref{s-flowdiagramKruskal}. We have fixed the initial value of $\phi$ to be a constant while plotting the curves. A portion of the Cauchy slice becomes null when 
\bea
s_{\text{null}} = e^\phi.
\eea 
For any $s > s_{\text{null}}$, the Cauchy surface hits the singularity and contains a timelike component. One should probably not take these evolutions seriously after they hit the singularity. It may be interesting to study this further. The BTZ singularity is in some ways a relatively trivial singularity compared to those in higher dimensional black holes, so it seems possible that one can study them more intensely from the boundary \cite{delaFuente:2013nba}. Note that a similar evolution of the Cauchy slices was observed in \cite{Leutheusser:2021frk} in the large mass limit. The crucial difference here is that we can directly derive these evolutions without requiring a large mass because our translations are point-wise, as opposed to the ones in \cite{Leutheusser:2021frk}.

\section{An Observer in the Interior}\label{infaller}

In our previous discussions, we have implicitly assumed that the state-independent operators are constructed via dressing to the boundary. In the two-sided eternal black hole, this can be understood as dressing to the left and right CFTs, while in the single-sided language, the ``simple'' and ``complicated'' operators both live ultimately in the right CFT. This is the usual picture of holography where the gravitational fluctuations die down at the boundary and one can construct (relational) observables via relating to the boundary.

In some recent papers \cite{Chandrasekaran:2022cip, Witten:2023qsv, Witten:2023xze}, Witten has argued that one can construct background independent algebras for quantum gravity by dressing instead to an observer's worldline. Similar thoughts (but without invoking the language of algebras) have appeared earlier in the context of gauge theories, see eg. \cite{AharonovSusskind}\footnote{Indeed, one can view these earlier ideas as crossed product constructions in a Type I setting. See some recent related discussions in \cite{Gomez}.}. The idea is to consider a Hamiltonian for the observer, and then to construct a crossed product via the total Hamiltonian of the system and the observer. This leads to non-trivial observables in gravitating theories where otherwise one would end up with a trivial algebra of observables \cite{Chandrasekaran:2022cip}.

In our setting, a particularly interesting application of this idea is to consider an observer in the interior. Note that a fairly generic timelike geodesic on either AdS wedge will eventually become an infaller, so this is a generic scenario. For discussions in this section, we will restrict our attention to the BTZ geometry -- some relevant results about the nature of timelike geodesic observers in eternal BTZ which will be important for us are derived in Appendix \ref{BTZApp}.

Following the philosophy outlined in \cite{Chandrasekaran:2022cip, Witten:2023qsv, Witten:2023xze}, perhaps the right way to think about an infaller or an interior observer is to imagine a quantum gravity theory, which contains a ``code subspace'' in bulk EFT that comprises of a black hole with such an observer. Because of the discussions in \cite{Leutheusser:2021frk, Witten:2021unn} and this paper, it is reasonable to think that such a code susbspace may exist in a holographic CFT. Interestingly, we will find that the results we obtain by adding an observer to Kruskal BTZ have natural interpretations in the holographic description. This is satisfying, because the discussions of \cite{Chandrasekaran:2022cip,Witten:2023qsv} were in settings where the UV theory is not known explicitly\footnote{Construction of a fully accepted de Sitter vacuum in string theory is an open problem, for example.}.


Following \cite{Witten:2023qsv} we will associate an algebra to the observer's worldline invoking the timelike tube theorem \cite{Strohmaier}. In order to do this, we need a Hamiltonian for the observer and the simplest choice is to set $H=q$ and to use $q$ and its canonically conjugate $p= -i \frac{d}{dq}$ as the operators associated to the observer. Note that with this choice, what is really accomplished in \cite{Witten:2023xze} is that we have an implementation of the energy-time commutation relations
\bea
[H,t]=i \hbar
\eea
where $H \equiv q$ and $t \equiv -p$ are operators. A similar philosophy, but with a more complicated implementation of the energy-time commutation relations, has appeared before, see eg. \cite{AharonovBohm2}. This construction essentially defines a clock for the observer.

\begin{figure}
\centering
\begin{subfigure}[b]{0.45\textwidth}
\centering
\includegraphics[width=\textwidth]{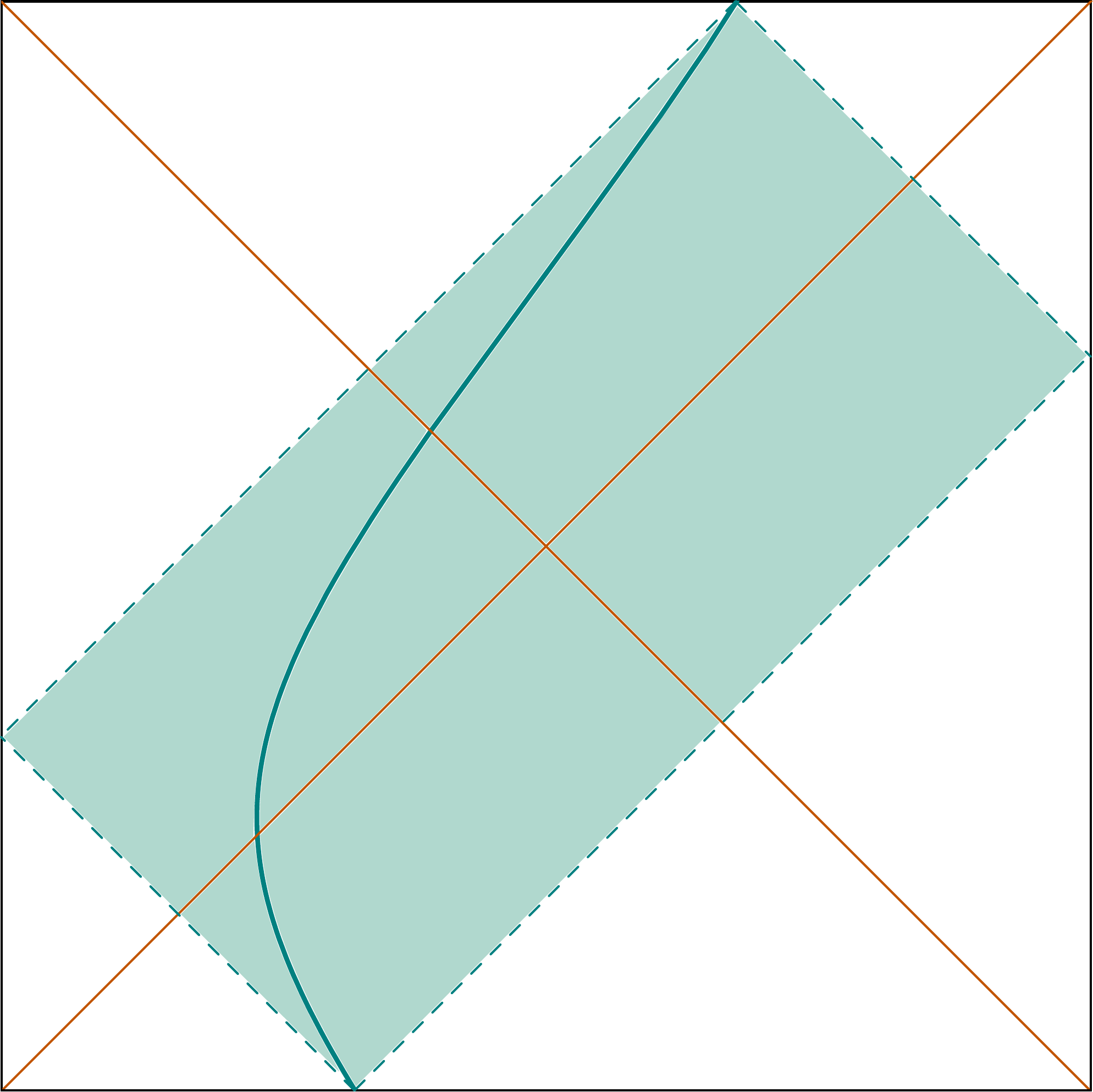}
\caption{}
\label{singularityobserverfig1}
\end{subfigure}
\hfill
\begin{subfigure}[b]{0.45\textwidth}
\centering
\includegraphics[width=\textwidth]{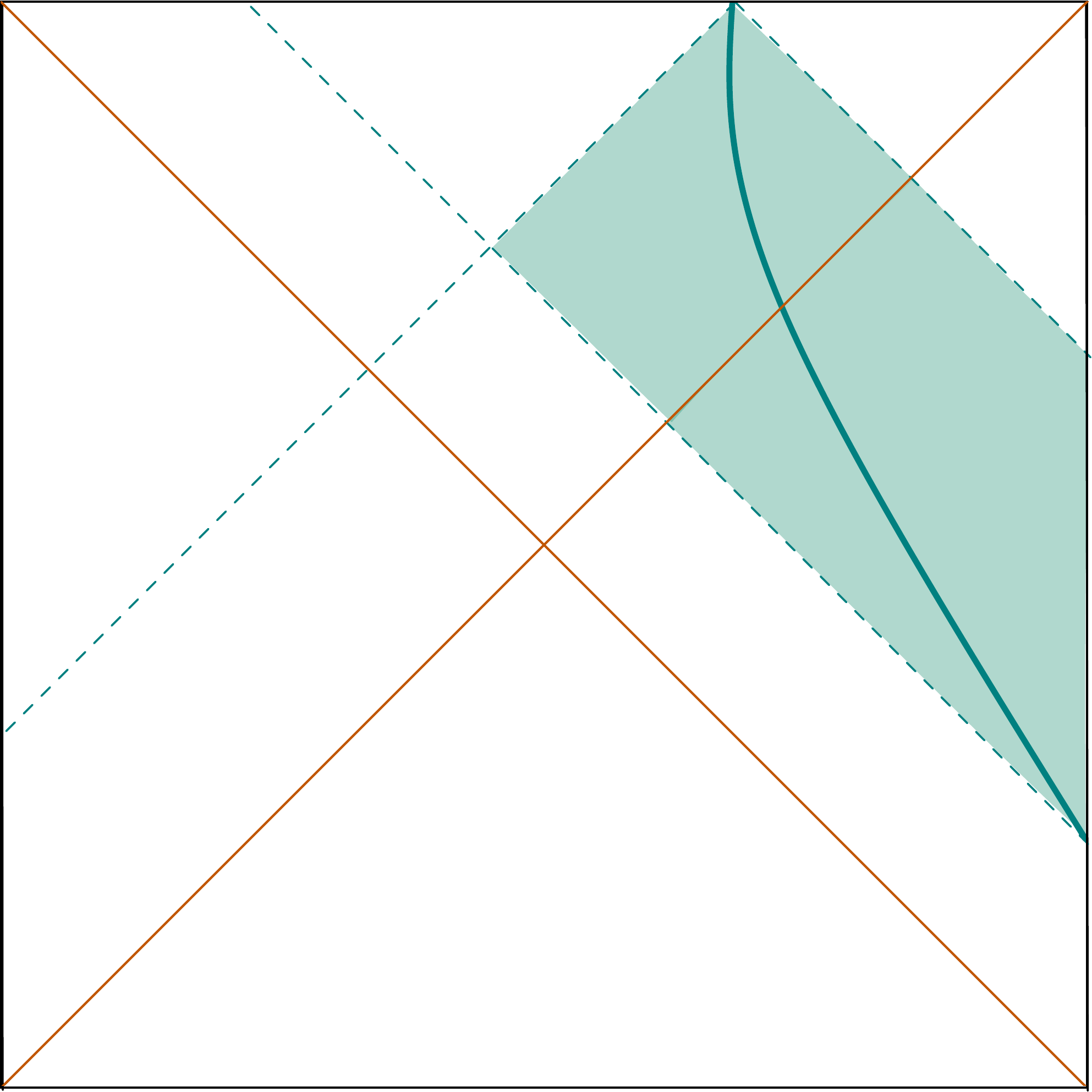}
\caption{}
\label{singularityobserverfig2}
\end{subfigure}
\caption{(a) The plot shows a geodesic that starts at the white hole singularity and ends in the black hole singularity. It is a general feature of such curves, irrespective of details that they have access to the entire bulk Cauchy slice. (b) Infalling geodesics have access only to partial Cauchy slices. Most curves of this kind look very close to null in actual plots of the Penrose diagram, so we have exaggerated the ``timelike-ness'' for clarity in this Figure. Figure (a) on the other hand, is an actual plot.}
\end{figure}

If we imagine that the observer's past light cone contains a complete Cauchy surface, then the algebra that is naturally associated is Type I \cite{Witten:2023xze}. It turns out that this is in fact the case for observers that start and end on the (future and past) singularities of the eternal BTZ black hole (See figure \ref{singularityobserverfig1}). The observer can access an entire spatial slice of the Kruskal manifold. We believe that the emergence of a Type I algebra in this case is natural also from a single-sided perspective. Indeed the mirror operator construction tells us that dynamics cannot be contained in the Type III$_1$ single trace sector alone after horizon-crossing, and acquires support in the ``complicated'' CFT operators as well. Therefore it is plausible that an observer that emerges from the white hole singularity, and ends in up in the black hole singularity would have access to the full algebra of CFT operators. The latter is indeed Type I. From the bulk, this manifests itself in the fact that the accessible Cauchy slice of such an observer is the entire Kruskal slice. Note that this is not (at least to us) obvious -- it is a geometric fact that timelike geodesics that start and end on (past and future) singularities have access to the entire Cauchy slice, which we prove in Appendix \ref{BTZApp}. The fact that Type I algebras have pure states may be an indication that the destruction of the observer at the singularity is part of the Type I description.

Another natural scenario is that of an observer whose proper time range is bounded in $t$ because of the singularity, who starts from one of the asymptotic boundaries. Such observers do not have access to the entire Cauchy slice, and therefore the algebra of observables is naively Type III. But one of the lessons from \cite{Chandrasekaran:2022cip, Witten:2023qsv, Witten:2023xze}, is that when there is an observer, we should do a crossed product construction with the total Hamiltonian $H_{Tot}=H_{sys} +H$ of the system and the observer to determine the correct algebra, $\mathcal{A}$. The resulting algebra is of type II because we choose $T$ to be $H_{sys}$, and the crossed product with this outer automorphism leads to type II (See \cite{Chandrasekaran:2022cip} for more details). A natural further demand on such constructions is that one should project the $\mathcal{A}$ algebra on to the algebra $\mathcal{A}'=\Theta(H) \mathcal{A} \Theta(H)$ to ensure that the observer Hamiltonian is bounded from below. The result is a Type II$_1$ algebra \cite{Chandrasekaran:2022cip, Witten:2023xze}. Note that the singularity played a crucial role in our argument above, because its presence was what allowed us to associate the original Type III algebra to the worldline of the observer -- the region accessible to the observer is a subregion of the bulk and therefore has a natural Type III algebra associated to it.


\section{Discussion}\label{disc}

The crossed product construction has been largely done at an abstract level in the recent literature, it will be interesting to make it more concrete. Essentially, the construction is about incorporating explicit dressings (anchored to boundaries or observers) to operators to re-define them. It seems likely that perturbative quantum field theory with these dressed operators instead of the bare operators will be better suited for incorporating perturbative gravity.

The fact that we could do a modular translation and then do the crossed product construction suggests that there may be a new translation operation, that is well-defined for Type II$_\infty$ algebras. A suitable operation of this kind would ensure that crossed producting and modular translations commute. At least in the case of generalized free field theories, we expect such crossed product compatible ``modular translations'' to exist for Type II$_\infty$ algebras. This may also be useful for making the state-independence of the construction, more manifest. Let us emphasize that these translations, if they exist, will not quite be the half-sided modular translations as the latter exist only for type III algebras. 

\subsection{The General Picture}\label{summary}

We conclude this paper by giving a qualitative summary of the overall picture that seems to be emerging from these results, together with some specific comments.

Our starting point in this paper was the realization of \cite{Leutheusser:2021frk, Leutheusser:2021qhd} that in the large-$N$ limit, the algebra of single trace operators around a thermofield double state above the Hawking-Page transition is type III$_1$. Note that at finite-$N$, the single trace algebra generates the full algebra of operators in the CFT and is therefore type I. Together with the observations of \cite{Witten:2021unn, CPW}, we take this to be a very strong suggestion that there is a specific large-$N$ limit where complexity is held fixed, in which heavy CFT states arrange themselves into backreacted TFD-like states and fluctuations around them. This latter perspective, is what is really doing most of the heavy-lifting for us. In such a limit, because $N$ has gone to infinity first, we never reach typicality under bulk time evolution. In other words, the perspective we have presented here following \cite{Leutheusser:2021frk, Leutheusser:2021qhd, Witten:2021unn, CPW} clarifies the notion of an ``equilibrium state'' in \cite{Papadodimas:2013jku} and fluctuations around it. We expect that even at finite-$N$ a smooth horizon will make sense (only?) to the extent that there is a trade off of this kind between complexity and $N$. This {\em emergent} nature of the black hole interior is obscured if one works directly at finite-$N$ and postulates a bulk Kruskal time.  

Naively, one might think that the interior reconstruction one does this way will depend on the cyclic separating vector. However, the crossed product construction shows us that when one incorporates perturbative $1/N^2$ corrections, the algebra of observables  in fact does {\em not} depend on this choice -- other states in the GNS Hilbert space (and not just the TFD state) will also lead to physically identical interior algebras. This removes the discomfort one feels when using mirror operators which apparently have an explicit dependence on the state. Stripping off technicalities, it means that in this limit, both the interior and exterior reconstructions have precisely the same degree of dependence on the (background) state\footnote{Note eg., that the fluctuation (ie., single trace) modes outside a black hole depend on the black hole, because they are obtained by solving wave equations.}.  It is worth noting that even for the exterior, the crossed product provides the natural way to think about state-independence \cite{Witten:2021unn}.  


Finally, let us comment on our perspective on the so-called ``frozen vacuum'' problem\footnote{We thank Suvrat Raju for raising this question.}. Frozen vacuum refers to the fact that if we were given a (heavy) CFT state and were to blindly use it to define mirror operators, then {\em by construction} we would conclude that the horizon is smooth for {\em any} such CFT state. This is problematic, because one expects from semi-classical bulk intuition that one can throw in small fluctuations into the horizon. These should look distinct from the vacuum at least for the duration of passage of these excitations through the horizon. This problem was addressed in \cite{Papadodimas:2013jku} by demanding that one choose a state $|\Psi\rangle$ to do the mirror construction first, and then in the excited state (which was implemented by the action of a unitary $U |\Psi\rangle$) one construct mirror operators after first ``undoing'' the $U$ via a conjugation. Let us first note that this has an eminently natural understanding in our picture -- the $|\Psi\rangle$ corresponds to the backreacted TFD-like state in the large-$N$ limit, while the fluctuations correspond to the single trace sector. The instruction of \cite{Papadodimas:2013jku} becomes the statement that we do reconstruction using the former, the background state\footnote{Note also that this is identical (and familiar) for the black hole exterior.}.  This apparent choice of state then is shown to be inconsequential at the level of the operator algebra by the Connes construction. We also learn something about the nature of allowed excitations. In the language of \cite{Papadodimas:2013jku} the unitary $U$ is constructed from the exponential of a single trace operator and therefore is not bounded in complexity. This is reasonable at infinite-$N$, but at finite-$N$ with the hindsight of our perspective built from \cite{Leutheusser:2021frk, Leutheusser:2022bgi, Witten:2021unn, CPW}, we expect that the (possibly approximately unitary) operator $U$ is better thought of as bounded in complexity, ie., constructed from at most $k$ single trace operators with $k \ll e^{\mathcal{O}(N)}$. 


\acknowledgments

We thank Aranya Bhattacharya, Justin David, Friðrik Freyr Gautason, Hong Liu, Suvrat Raju and Lárus Thorlacius for discussions/correspondence. We also thank Hong Liu and Lárus Thorlacius for their comments on the draft. We thank the JHEP referee for numerous comments which helped improve the manuscript, in particular for suggesting the inclusion of Section 1.1 as well as catching an error in the previous Figure 3(b). We thank Antony Speranza whose comments have improved the discussion in sections 3 and 5 of arXiv V3. VM is supported by the Icelandic Research Fund under grant 228952-052.

\appendix

\section{Crossed Product Alegbras}
\label{crossedproductrevsec}

Let us review the crossed product construction in \cite{AliAhmad:2023etg,Witten:2021unn}. Consider a von Neumann algebra $\mathcal{U}$. We will assume that the algebra has a representation on a Hilbert space $\mathcal{H}$. If $T$ generates a group of automorphisms, then we have
\bea
e^{isT}ae^{-isT} \in \mathcal{U} \quad \quad \forall a \in \mathcal{U}, s \in \mathbb{R}.
\eea
If $e^{isT} \in \mathcal{U}$, we will call the automorphism an inner automorphism. Otherwise, we will call it an outer automorphism. Let $X$ be some bounded function in $\mathbb{R}$. Then, the crossed product algebra
\bea
\hat{\mathcal{U}} = \mathcal{U} \rtimes \mathbb{R}
\eea
acting on the Hilbert space
\bea
\hat{\mathcal{H}} =\mathcal{H} \otimes L^{2}(\mathbb{R}).
\eea
is given by adjoining $T+X$ to $\mathcal{U}$. The elements of the algebra are then given by
\bea
ae^{isT}\otimes e^{isX} \quad \quad s\in \mathbb{R}, a \in \mathcal{U}.
\eea
Let us assume that the algebra $\mathcal{U}$ is of type III$_1$. If $T$ is an outer automorphism, then the crossed product algebra $\hat{\mathcal{U}}$ is of type II$_{\infty}$ \cite{takesaki1973duality}.

If the algebra has a cyclic, separating vector $\Psi \in \mathcal{H}$, we can define a modular Hamiltonian associated to the vector. Tomita-Takesaki theory says that the modular Hamiltonian generates automorphisms of $\mathcal{U}$. In \cite{Witten:2021unn}, the outer automorphism generator was chosen as the modular Hamiltonian $H$. Moreover, the operator $X$ was chosen to be $H_L$\footnote{In \cite{Chandrasekaran:2022cip}, $X$ was chosen to be the Hamiltonian of an observer in the static patch of dS.}. Therefore, adjoining $T+X$ is equivalent to adjoining $H_R$. Crucially, the crossed product $\hat{\mathcal{U}}$ is independent of the vector $\Psi$.

\section{Extended Half-Sided Modular Translations}
\label{modulartransappendix}

Consider a von Neumann algebra $\mathcal{M}$ acting on a Hilbert space $\mathcal{H}$. Let $\mathcal{N}\subset \mathcal{M}$ be a subalgebra of $\mathcal{M}$. If $\Omega$ is a common cyclic and separating vector for both $\mathcal{M}$ and $\mathcal{N}$, then we associate the operators $\Delta_{\mathcal{M},\mathcal{N}}$ and $J_{\mathcal{M},\mathcal{N}}$ to the algebras. Let us also assume that $\mathcal{N}$ is preserved under the modular flow of $\Delta_{\mathcal{M}}$:
\bea
\Delta_{\mathcal{M}}^{i t} \mathcal{N} \Delta_{\mathcal{M}}^{-i t} \subset \mathcal{N} \quad \forall t \leq 0 \label{modularinclusion}
\eea
If these conditions are met, there exists a unitary operator $U(s), s \in \mathbb{R}$ satisfying the properties \cite{Borchers:1991xk,Wiesbrock:1992mg,10.1063/1.533323,Leutheusser:2021frk}:
\begin{enumerate}
\item $U(s)$ is generated by a bounded operator:
\bea
U(s) = e^{-iGs}, \quad \quad G\geq 0. \label{uspointseq1}
\eea
\item $U(s)$ leaves the cyclic separating vector invariant
\bea
U(s)\Omega =\Omega \quad \forall s \in \mathbb{R}
\eea
\item $U(s)$ is a half-sided inclusion
\bea
U^{\dagger}(s)\mathcal{M}U(s) \subseteq \mathcal{M} \quad s\leq0
\eea
\item The subalgebra $\mathcal{N}$ can be mapped to $\mathcal{M}$ through the relation
\bea
\mathcal{N} = U^{\dagger}(-1)\mathcal{M}U(-1)
\eea
\item $\Delta_{\mathcal{M}}$ and $J_\mathcal{M}$ act on $U(s)$ acts
\bea
\Delta_{\mathcal{M}}^{i t} U(s) \Delta_{\mathcal{M}}^{-i t} = U(e^{-2\pi t}s) \quad \quad J_\mathcal{M} U(s) J_\mathcal{M} = U(-s) \quad \forall t,s \in \mathbb{R}. \label{deltauconjugation}
\eea
\item U(s) forms a 1-parameter family of unitary automorphism group:
\bea
U(s)^{\dagger} =U(-s), \quad \quad U(s_1)U(s_2) = U(s_1+s_2). \label{uspointseq2}
\eea
\end{enumerate}
The resulting structure is referred to as a half-sided modular translation. It can be shown that if $\mathcal{M}$ is a factor, then $\mathcal{M}$ has to be of type III$_1$ \cite{Wiesbrock:1992mg}. We can also define half-sided modular inclusions \eqref{modularinclusion} for $t\geq0$. This gives us the half-sided modular translations for $s\geq0$. These translations will satisfy all the above properties, provided we flip the sign of $t$ on the right hand side of the first equation in \eqref{deltauconjugation}.

Now, let us look at generalized free field theories. It was shown in \cite{Leutheusser:2021frk} that the general structure of $U(s)$ can be determined for all values of $s$ without specifying $\mathcal{N}$. To see this, let us look at a scalar field $\phi_R$ acting on the right exterior wedge of a black hole spacetime. The evolution of this operator under $U(s)$ is given by
\bea
\phi_R(X;s) = U^{\dagger}(s)\phi_R(X)U(s)
\eea
As the field can be expressed in terms of oscillators $a_k$ as described in equation \eqref{fieldexpansioneq}, we can infer the evolution of $\phi_R$ by examining how $U(s)$ acts on the oscillators. We have
\bea
U^{\dagger}(s)a_k^{(\alpha)}U(s) = \Lambda^{\alpha\beta}_{kk^{\prime}}a_k^{(\beta)} \quad \quad \alpha,\beta =R,L
\eea
where $a_k^{(R)}$ and $a_k^{(L)}$ are the oscillators of the right and the left wedge. In our discussion, it is implied that repeated indices should be summed over. Using the properties of the half-sided modular translations, we can show that \cite{Leutheusser:2021frk}
\bea
\Lambda_{k^{\prime} k}^{R R}(s)=\left\{\begin{array}{ll}
C_{k^{\prime} k}(s) & s<0 \\
\frac{\sinh \pi \omega^{\prime}}{\sinh \pi \omega} C_{k^{\prime} k}(-s) & s>0
\end{array} \quad, \quad \Lambda_{k^{\prime} k}^{R L}(s)=\left\{\begin{array}{ll}
0 & s<0 \\
\frac{\sinh \pi\left(\omega+\omega^{\prime}\right)}{\sinh \pi \omega} C_{k^{\prime}-k}(-s) & s>0
\end{array},\right.\right.\nonumber
\eea
\bea
\Lambda_{k^{\prime} k}^{L L}(s)=\left\{\begin{array}{ll}
\frac{\sinh \pi \omega^{\prime}}{\sinh \pi \omega} C_{-k^{\prime}-k}(s) & s<0 \\
C_{-k^{\prime}-k}(-s) & s>0
\end{array} \quad, \quad \Lambda_{k^{\prime} k}^{L R}(s)=\left\{\begin{array}{ll}
\frac{\sinh \pi\left(\omega+\omega^{\prime}\right)}{\sinh \pi \omega} C_{-k^{\prime} k}(s) & s<0 \\
0 & s>0
\end{array} .\right.\right.\nonumber\\
\eea
where
\bea
C_{k k^{\prime}}(s)=\sqrt{\frac{\sinh \pi\left|\omega^{\prime}\right|}{\sinh \pi|\omega|}} e^{i \gamma_k-i \gamma_{k^{\prime}}} I_{\omega \omega^{\prime}}(s) \delta_{q q^{\prime}},\quad \quad I_{\omega \omega^{\prime}}(x) = x^{-i(\omega- \omega^{\prime})}\Gamma\left(\omega- \omega^{\prime}+i\epsilon\right).\nonumber\\
\eea
The action of $U(s)$ on the field is completely specified up to a phase factor $e^{i\gamma_k}$. This leftover degree of freedom corresponds to the choice of the subalgebra $\mathcal{N}$.

\section{Tomita Operator of the Shifted Wedge}\label{Tomita}

The Tomita operator $S$ associated to the thermofield state $\Psi_{\text{TFD}}$ is given by \cite{WittenEntanglement}
\bea
S a \Psi_{\text{TFD}} = a^{\dagger} \Psi_{\text{TFD}}, \quad \text{where} \ a \in \mathcal{A}
\eea
The modular operator $\Delta$ and the modular conjugation $J$ are related to the operator through the relations
\bea
\Delta = S^{\dagger}S \quad \quad J = S\Delta^{-\frac{1}{2}}. \label{modularopandconjeq}
\eea
We can construct the Tomita operator of the shifted wedge, which we will denote as $S^{\prime}$, as follows
\bea
\tilde{S} = U^{\dagger}(s_0)S U(s_0) \label{newtomita}
\eea
For some $\tilde{a} \in \mathcal{U}$, we can see that
\bea
\begin{aligned}
\tilde{S} \tilde{a} \tilde{\Psi}_{\text{TFD}} &= \left(U^{\dagger}(s_0)S U(s_0)\right)\left(U^{\dagger}(s_0)a U(s_0)\right)U^{\dagger}(s_0)\Psi_{\text{TFD}} \\
&= \left(U^{\dagger}(s_0)a^{\dagger} U(s_0)\right)U^{\dagger}(s_0)\Psi_{\text{TFD}} \\
&=\tilde{S} \tilde{a}^{\dagger} \tilde{\Psi}_{\text{TFD}}
\end{aligned}
\eea
as expected. Using \eqref{modularopandconjeq} and \eqref{newtomita}, we can define new modular operator $\tilde{\Delta}$ and modular conjugation $\tilde{J}$. Since $U(s_0)$ is a unitary operator, the expressions simplify and we get \cite{Lashkari:2018oke}
\bea
\tilde{U} = U^{\dagger}(s_0)S U(s_0), \quad \quad \tilde{J} = U^{\dagger}(s_0)J U(s_0).
\eea

\section{Eternally Ill-fated Observers}\label{BTZApp}

In this section, we will look at the geodesics of observers that start at the past singularity and end at the future singularity. We will work in the Kruskal BTZ metric in \eqref{Kruskalmetric}. We will show that the accessible region to these observers is the entire Cauchy slice of the Kruskal manifold. We suspect this observation, or a closely related variation, may be true in general Kruskal geometries.

The metric is invariant under the transformations $u\to e^{-\lambda}u$ and $v\to e^{\lambda}v$. We can see that these transformations correspond to the usual time-translation symmetry in the exterior global static coordinates. The Killing vector corresponding to this symmetry can be immediately read off as follows:
\bea
K = K^{\mu}\partial_{\mu} = -u \partial_{u}+v \partial_{v}
\eea
Inverting the components, we get
\bea
K_{\mu} = \left(\frac{-4 \ell^{2}v}{(1+u v)^{2}}, \frac{4 \ell^{2}u}{(1+u v)^{2}},0\right)
\eea
Along a geodesic, we know that $K_\mu \frac{dx^{\mu}}{d\tau}$ is a constant. This gives us the following conserved quantity:
\bea
\frac{4 \ell^{2}}{(1+u v)^{2}} \left(v\dot{u}-u\dot{v}\right) = E \label{geoeq1}
\eea
Using the rotational invariance of the metric, we can also get another conserved quantity:
\bea
R^{2}\frac{(1-u v)^{2} }{(1+u v)^{2}} \dot{\phi} = L
\eea
Since we are looking at timelike geodesics, we choose the parametrization of the curves by imposing
\bea
\frac{-4 \ell^{2} }{(1+u v)^{2}} \dot{u} \dot{v} +R^{2}\frac{(1-u v)^{2} }{(1+u v)^{2}} \dot{\phi}^{2} = -1 \label{kruskalnormeq}
\eea
This gives us three coupled first-order differential equations in three variables. We will set $\dot{\phi}=0$ to simplify the discussion. Solving the remaining equations, we obtain
\bea
v(u)=\frac{C_{2}+\left(C_{1}+C_{2}^{2}\right) u}{1+C_{2} u} \label{kruskalgeodesicsoleq}
\eea
where $C_{1,2}$ are constants of integration. Let us look at the geodesics that start (end) at the past (future) singularity, given by the curve $uv =1$. At the singularity, we have
\bea
\frac{C_{2}+\left(C_{1}+C_{2}^{2}\right) u}{1+C_{2} u} = \frac{1}{u} \quad \implies \quad u = \pm\frac{1}{\sqrt{C_{1}+C_{2}^{2}}}.
\eea
Therefore, the endpoints of the geodesics, given by
\bea
(u,v) = \left(\pm\frac{1}{\sqrt{C_{1}+C_{2}^{2}}},\pm\sqrt{C_{1}+C_{2}^{2}}\right),
\eea
are symmetric w.r.t the horizons. This leads to interesting consequences. In particular, let us look at the region casually accessible to the observer. The overlap of the past lightcone of the future endpoint and the future lightcone of the past endpoint gives this region. Using straightforward trigonometric arguments, it is easy to see that the causally accessible region of these observers forms a rectangle in the Penrose diagram, with two corners on the asymptotic boundary and the other two at the singularities.

\bibliographystyle{JHEP}
\bibliography{refs}

\end{document}